\documentclass[%
 reprint, 
unsortedaddress,
 amsmath,amssymb,
 aps, physrev,
floatfix,
]{revtex4-2}

\usepackage{mathsfbf}
\usepackage{graphicx}
\usepackage{dcolumn}
\usepackage{bm}
\usepackage{hyperref}
\usepackage[mathlines]{lineno}
\usepackage{xcolor}

\begin{document}


\title{\textbf{Modelling bulk mechanical effects in a planar cellular monolayer} 
}%

\author{Natasha Cowley}
 \email{Natasha.Cowley@manchester.ac.uk}
\affiliation{Department of Mathematics, University of Manchester, Oxford Road, Manchester M13 9PL, UK}

\author{Sarah Woolner}
\email{Sarah.Woolner@manchester.ac.uk}
\affiliation{Division of Cell Matrix Biology \& Regenerative Medicine, Unviersity of Manchester, Oxford Road, Manchester, M13 9PT, UK}

\author{Oliver E. Jensen}
\email{Oliver.Jensen@manchester.ac.uk}
\affiliation{Department of Mathematics, University of Manchester, Oxford Road, Manchester M13 9PL, UK}

\date{\today}

\begin{abstract}
We use a three-dimensional formulation of the cell vertex model to describe the mechanical properties of a confluent planar monolayer of prismatic cells.  Treating cell height as a degree of freedom, we reduce the model to a two-dimensional form.  We show how bulk effects, associated with cell volume and total surface area, lead to coupling between energy variations arising from changes in cell apical area and apical perimeter, a feature missing from standard implementations of the two-dimensional vertex model.  The model identifies {\color{black}five} independent mechanisms by which cells can lose in-plane rigidity, relating to variations in total cell surface area, the strength of lateral adhesion, and constrictive forces at the apical cortex.  The model distinguishes bulk from in-plane stresses, and identifies two primary measures of cell shear stress.  In the rigid regime, the model shows how lateral crowding in a disordered isolated monolayer can lead to cell elongation towards the monolayer centre.  We examine loss of in-plane rigidity in a disordered monolayer and connect isolated patches of stiffness that persist during the rigidity transition to the spectrum of a Laplacian matrix.  This approach enables bulk mechanical effects in an epithelium to be captured within a two-dimensional framework.
\end{abstract}

\maketitle

\section{Introduction}

Given their sheet-like structure, it is common for epithelia to be modelled using two-dimensional (2D) formulations.  The cell vertex model is a widely used framework that generates realistic cell patterns and offers insights into in-plane stresses, making it a popular tool with which to examine how epithelial cells generate and respond to mechanical forces \cite{farhadifar2007}.  Here, we show how three-dimensional (3D) effects can be incorporated systematically within a 2D framework, revealing relationships between 2D (in-plane) and 3D (bulk) stresses.  

Unlike traditional hyperelasticity, in which mechanical energy is described in terms of strains, the vertex model assigns an energy to a multicelluar tissue that depends on geometric invariants (lengths, areas and volumes of individual cells).  The vertex model is now routinely implemented in order to investigate the organisation of tissues in 3D \cite{gomez2021complex}, facilitated by the release of computational tools \cite{sego2023general, runser2023simucell3d}.   In comparison to its standard 2D formulation, in which energy is typically assumed to be the sum of convex functions of cell apical area and apical perimeter, 3D vertex models offer a broader range of constitutive choices.  The energy may include a contribution that is linear in contact area between adherent cells {\color{black}\cite{honda2004three, okuda2013, drozdowski2024morphological}}, linear in the area exposed to surroundings {\color{black}\cite{honda2004three, okuda2013, hannezo2014}}, linear in the area between heterotypic cells \cite{sahu2021geometric, villeneuve2024mechanical}, quadratic in cell volume relative to a target value $V_0^*$ {\color{black}\cite{honda2004three, okuda2013, merkel2018geometrically}}, quadratic in total cell surface area relative to a target area $a_0^*$ \cite{merkel2018geometrically}, or quadratic in cell edge lengths \cite{ioannou2020development}; see \cite{khan2023single} for an overview of some of the different combinations assumed by prior authors. Additional energy contributions may penalise bending of cell edges \cite{durney2021three} or bending of adjacent cells within a sheet \cite{du2014}.

Beyond providing the opportunity to infer stresses in an epithelium from primarily geometric information \cite{noll2020, ogita2022}, the 2D vertex model has attracted considerable attention as a tool with which to investigate rigidity transitions in planar monolayers \cite{bi2015}.  In the standard 2D framework, rigidity of individual cells arises {\color{black}typically} from a balance between forces that resist changes in cell apical area and forces that resist changes in the apical perimeter.  The latter forces regulate resistance to in-plane shear, enabling the rigidity of a monolayer to be regulated through changes of a single dimensionless parameter.  A rigidity transition in a 3D multicellular tissue with energy defined in terms of cell volume and total surface area has been shown to be regulated by the parameter $a_0^*/V_0^{*2/3}$ \cite{merkel2018geometrically, zhang2022topologically, kim2024mean}.  In the configuration that we investigate below, using a 3D description of a planar monolayer, we identify {\color{black}five} independent dimensionless parameters that may regulate the in-plane rigidity transition, broadening the potential repertoire of mechanisms by which cells may exploit this change of phase. 

The vertex model is also a valuable tool with which to understand the mechanisms by which epithelial cells in a monolayer regulate their shape.  The transition between columnar (elongated), cuboidal and squamous (flat) cell shapes will be regulated by competition between multiple competing effects \cite{mao2015, kondo2015}.  Shape changes may be driven by molecular pathways, such as the cuboidal/squamous transition in \textit{Drosophila} ovary epithelium that is regulated by Tao, which endocytoses the cell-cell adhesion molecule Fasciclin-2 \cite{gomez2012}, or in \textit{Drosophila} wing where a columnar/cuboidal transition is driven by cortical actomyosin contractility \cite{ray2018}.  Alternatively, cell shape may change as a result of external stretching forces, for example in the urothelium during filling of the bladder \cite{khandelwal2009} or during bending of the \textit{Drosophila} wing disc \cite{harmansa2024}.  Proliferation can be expected to promote cell crowding \cite{mao2015}, which may influence tissue stiffness \cite{chisolm2025} and drive cells to adopt complex 3D shapes \cite{barone2024}.  This motivates investigation of simple mechanical models that capture the dominant physical balances that determine cell shape. 

In 2D formulations, it is common to use `pressure' to describe energy changes with respect to cell apical area, and `tension' to describe energy changes with respect to cell apical perimeter.  The terminology can be confusing, given that (by resisting area changes) the 2D `pressure' is analogous to the surface tension of a liquid.  In 3D, we will follow convention by referring to energy changes with respect to cell volume as pressure, and changes with respect to perimeter as tension, while both terms are used interchangeably to describe changes with respect to area.  To add further complication, different sign conventions are adopted by different authors when defining pressure; we avoid introducing sign differences between pressures and tensions when defining them as energy derivatives.

\begin{figure}[t]
    \centering
    \includegraphics[width=0.75\linewidth]{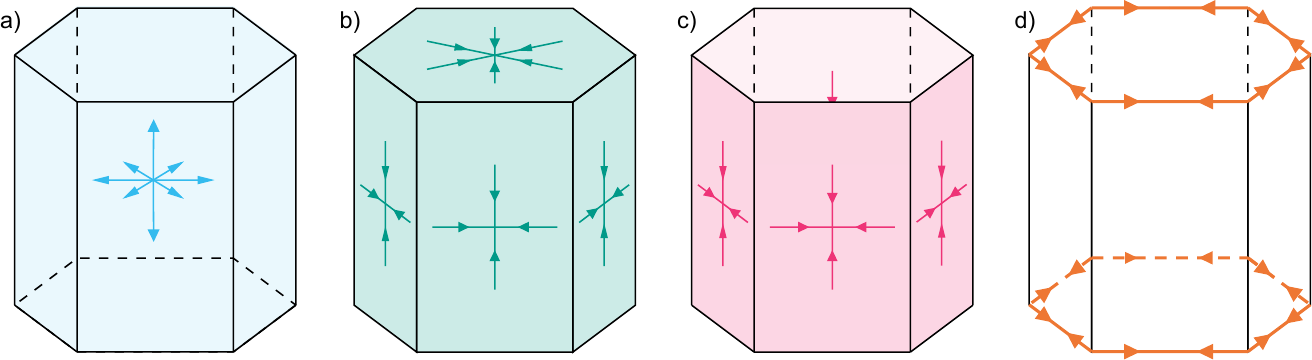}
    \caption{A schematic illustrating the four contributions to the mechanical energy of a prismatic cell, associated with changes of (a) volume, (b) total surface area, (c) lateral surface area and (d) apical perimeter (with assumed apical-basal symmetry).}
    \label{fig:schematic}
\end{figure}

In this study, we consider a planar monolayer of cells of uniform (or near-uniform) thickness, seeking to derive a 2D model that retains 3D mechanical features. 
To facilitate this, we impose apical-basal symmetry, postponing investigation of more complex structural features such as scutoids \cite{gomez2018scutoids, rozman2024basolateral} and cell nuclei. We pose an energy that is quadratic in cell volume, total area, apical and basal perimeter, and linear in adherent area between neighbouring cells (features illustrated in Fig.~\ref{fig:schematic}).  Imposing a constraint that restricts height variations between neighbouring cells, we use a force balance to determine the thickness of the cell monolayer (in contrast to others {\color{black}\cite{rozman2024basolateral, sarkar2025}} who prescribe the thickness).  We disregard curvature of the monolayer, recognising its importance in many applications \cite{deMarzio2025, harmand2021}, in order to focus on the implications of our chosen constitutive model. Our approach is similar in spirit to \cite{hannezo2014}, \cite{chiang2024} {\color{black}and \cite{sarkar2025}}, who also reduced 3D models of epithelial sheets to 2D formulations: we disregard cell-substrate adhesion \cite{hannezo2014} and we do not adopt a phase-field approach \cite{chiang2024}; {\color{black} but like \cite{hannezo2014} and \cite{sarkar2025}, we recover a sharp squamous/columnar transition in some circumstances}.  We compare the reduced 2D model to the traditional 2D vertex model in which apical area and apical perimeter contributions make independent contributions to the total energy; despite having {\color{black}five} independent dimensionless parameters at our disposal, we find no direct reduction that leads to the classical model {\color{black}unless cell height is prescribed}.  We examine rigidity transitions as functions of 3D parameters for a monolayer of identical hexagonal prisms.  For monolayers that are spatially disordered, we investigate relationships between 2D and 3D stress components, discuss rigidity in terms of connected regions of stiffness, and demonstrate crowding-induced elongation of cells at the centre of an isolated monolayer in the rigid regime.  The model is laid out in Sec.~\ref{sec:model} and numerical results are presented in Sec.~\ref{sec:results}.

\section{Model and methods}
\label{sec:model}

\subsection{Prismatic cells}

We consider a planar monolayer formed by $N_c$ confluent polyhedral cells (Fig.~\ref{fig:schematic}).  The cells are initially assumed to be confined between horizontal planes $z=0$ (basal) and $z=H$ (apical).  Cell shapes are defined by vertices lying in each plane.  We assume the cells are prismatic, meaning that each cell has equal numbers of apical and basal vertices.  We use a mechanical energy, defined in terms of 3D cell shapes, to move vertices horizontally towards an equilibrium configuration, while imposing a global vertical force balance in order to determine the cells' height $H$.  These restrictions allow us to recover a 2D formulation that incorporates some significant 3D geometric effects.

In what follows, we index cells with $h=1,\dots, N_c$, faces with $i=1,\dots,N_f$, edges with $j=1,\dots,N_e$ and vertices with $k=1,\dots,2N_v$.  The normals assigned to apical and basal faces of the monolayer are assumed to be parallel to the constant vector $\hat{\mathbf{z}}$.  Cell shapes are specified in terms of vertex locations $\mathbf{R}^\pm_k\in \mathbb{R}^3$ and the depth $H$.  We assume vertices $\mathbf{R}_k^-$, for $k=N_v+1,\dots,2N_v$ lie on the plane $z=0$ and vertices $\mathbf{R}_k^+$, for $k=1,\dots,N_v$, lie on the plane $z=H$.  Edges and faces are indexed in the order apical ($+$), lateral ($l$) and basal ($-$), so that $N_e=N_e^++N_e^l+N_e^-$ and $N_f=N_f^++N_f^l+N_f^-$.  Writing $\mathbf{R}_k^-=\mathbf{r}_k^-$ for $k=N_v+1,\dots,2N_v$ and $\mathbf{R}_k^+=\mathbf{r}_k^+ +H\hat{\mathbf{z}}$ for $k=1,\dots,N_v$, where $\hat{\mathbf{z}}\cdot\mathbf{r}_k^\pm\equiv 0$, we evaluate geometric quantities in terms of vertex locations $\mathbf{r}_k^\pm$, expressing these quantities in terms of signed incidence and adjacency matrices (see Appendix~\ref{app:matrices}).

We gather in-plane vertex locations into $\mathsfbf{r}^+\equiv (\mathbf{r}_1^+,\dots,\mathbf{r}_{N_v}^+)^\top$ and $\mathsfbf{r}^-\equiv (\mathbf{r}_{N_v+1}^-,\dots,\mathbf{r}_{2N_v}^-)^\top$.  Then, for cell $h$, we evaluate the cell volume $V_h(\mathsfbf{r}^+,\mathsfbf{r}^-,H)$, total cell area $a_h=a_h(\mathsfbf{r}^+,\mathsfbf{r}^-,H)$, apical and basal areas and perimeters $A_h^\pm(\mathsfbf{r}^\pm)$ and $L_h^\pm(\mathsfbf{r}^\pm)$, and lateral face areas $A_i^l(\mathsfbf{r}^+,\mathsfbf{r}^-,H)$.  We then assume that the mechanical energy of the system can be written as 
\begin{equation}
U=U(\mathsf{V}, \mathsf{a}, \mathsf{L}^+, \mathsf{L}^-, \mathsf{A}^l), 
\label{eq:genenergy}
\end{equation}
where $\mathsf{V}=(V_1,\dots,V_{N_c})$, $\mathsf{a}=(a_1,\dots,a_{Nc})$, $\mathsf{L}^\pm=(L_1^\pm,\dots,L^\pm_{N_c})$, $\mathsf{A}^l=(A_{N_f^++1}^l,\dots,A^l_{N_f^++N_f^l})$.  This formulation captures dependence on acto-myosin cortical rings via $\mathsf{L}^\pm$ and cell-cell adhesion via $\mathsf{A}^l$.  From this energy, we define respectively the bulk pressure $\mathsf{P}$, surface tension (or surface pressure) $\mathsf{T}$, cortical line tension $\mathsf{T}^{p\pm}$ acting over the apical and basal perimeters, and the lateral surface tension $\mathsf{T}^l$ related to adhesion.  These discrete fields have elements
\begin{subequations}
\label{eq:pderivs}
\begin{align}
    P_h&=\frac{\partial U}{\partial V_h}, \quad
    T_h=\frac{\partial U}{\partial a_h}, \quad
    T_h^{p\pm}=\frac{\partial U}{\partial L_h^\pm},\\
    T^l_i&=\frac{\partial U}{\partial A_i^l},
\end{align}
\end{subequations}
for $h=1,\dots, N_c$ and $i-N_f^+=1,\dots,N_f^l$.  Then, noting that $\mathsfbf{r}^+$, $\mathsfbf{r}^-$ and $H$ are independent degrees of freedom ($\mathsfbf{r}^\pm$ move only in a plane orthogonal to $\hat{\mathbf{z}}$), we can evaluate the energy change arising from virtual displacements of the vertices and a change in cell depth as
\begin{widetext}
\begin{align}
\mathrm{d}U=&{\sum_{k=1}^{N_v}} \left\{\left({\textstyle\sum_h} \left\{ 
P_h\frac{\partial V_h}{\partial \mathbf{r}_k^+}
+T_h\frac{\partial a_h}{\partial \mathbf{r}_k^+}
+T_h^{p+}\frac{\partial L_h^+}{\partial \mathbf{r}_k^+}\right\}
+{\textstyle\sum_i} T^l_i \frac{\partial A_i^l}{\partial \mathbf{r}_k^+}
\right)\cdot \mathrm{d}\mathbf{r}_k^+ \right\}
+\left({\textstyle\sum_h} \left\{
P_h\frac{\partial V_h}{\partial H}
+T_h\frac{\partial a_h}{\partial H}\right\}
+{\textstyle\sum_i} T^l_i \frac{\partial A_i^l}{\partial H}
\right)\mathrm{d}H
\nonumber \\
&+\sum_{k=N_v+1}^{2N_v} \left\{\left({\textstyle\sum_h} \left\{
P_h\frac{\partial V_h}{\partial \mathbf{r}_k^-}
+T_h\frac{\partial a_h}{\partial \mathbf{r}_k^-}
+T_h^{p-}\frac{\partial L_h^-}{\partial \mathbf{r}_k^-}\right\}
+{\textstyle\sum_i} T^l_i \frac{\partial A_i^l}{\partial \mathbf{r}_k^-}
\right)\cdot \mathrm{d}\mathbf{r}_k^- \right\}.
\label{eq:du}
\end{align}
A dynamic implementation of the cell vertex model involves $2N_v+1$ ordinary differential equations (ODEs) describing the evolution of $\mathsfbf{r}^\pm(t)$ and $H(t)$ down the energy gradients identified in (\ref{eq:du}) to an equilibrium state. 

We {\color{black}then} assume apical-basal symmetry, allowing us to write $\mathbf{r}_k^+=\mathbf{r}_{k+N_v}^-\equiv\mathbf{r}_k$, where $\mathbf{r}_k$ is a vector confined to a plane orthogonal to $\hat{\mathbf{z}}$. This simplifies the geometric quantities substantially, allowing them to be expressed in terms of apical areas and perimeters as $V_h=V_h(A_h^+,H)$, $a_h=a_h(A_h^+,L_h^+,H)$ and $A^l_i=A^l_i(t_i^+,H)$, where $t_i^+$ identifies the apical edge associated with lateral face $i$.   Specifically, 
\begin{subequations}
    \label{eq:geo_quants}
\begin{align}
    V_h&=H A_h^+,  & a_h&=2 A_h^++H L_h^+, & (h=1,\dots, N_c),  \label{eq:va} \\
A_i^l&=Ht_i^+,  & t_i^+&\equiv {\textstyle{\sum_j}} \vert B_{ij}^{l+}\vert \vert\mathbf{t}^+\vert_j, &(i=N_f^++1,\dots,N_f^++N_f^l).
\label{eq:aa}
\end{align}
\end{subequations}
(Here $\mathsf{B}^{l+}$ is a block of the edge-face incidence matrix; see Appendix~\ref{app:matrices}; geometric relationships are summarised in Appendix~\ref{app:geom}).  Treating $\mathsfbf{r}^+$ and $H$ as the $N_v+1$ degrees of freedom of the model, noting that $A_h^+=A_h^+(\mathsfbf{r}^+)$, $L_h^+=L_h^+(\mathsfbf{r}^+)$ and $t_i^+=t_i^+(\mathsfbf{r}^+)$, it follows from (\ref{eq:du}, \ref{eq:geo_quants}) that
\begin{align}
\mathrm{d}U=&2\sum_{k=1}^{N_v}\left({\textstyle\sum_h} \left\{ 
P_h\frac{\partial V_h}{\partial \mathbf{r}_k^+}
+T_h\frac{\partial a_h}{\partial \mathbf{r}_k^+}
+T_h^{p+}\frac{\partial L_h^+}{\partial \mathbf{r}_k^+}\right\}
+{\textstyle\sum_i} T_i^l \frac{\partial A^l_i}{\partial \mathbf{r}_k^+}
\right)\cdot \mathrm{d}\mathbf{r}_k^+ 
+\left({\textstyle\sum_h} \left\{
P_h\frac{\partial V_h}{\partial H}
+T_h\frac{\partial a_h}{\partial H}\right\}
+{\textstyle\sum_i} T_i^l \frac{\partial A^l_i}{\partial H}
\right)\mathrm{d}H \nonumber \\
=&2\sum_{k=1}^{N_v}\left({\textstyle\sum_h} \left\{ 
(P_h H +2T_h) \frac{\partial A_h^+}{\partial \mathbf{r}_k^+}
+(T_h H +T_h^{p+}) \frac{\partial L_h^+}{\partial \mathbf{r}_k^+}\right\}
+{\textstyle\sum_i} T_i^l H \frac{\partial t_i^+}{\partial \mathbf{r}_k^+}
\right)\cdot \mathrm{d}\mathbf{r}_k^+ \nonumber \\
&\qquad +\left({\textstyle\sum_h} \left\{P_h A_h^+ +T_h L_h^+\right\}+{\textstyle\sum_i} T_i^l t_i^+
\right)\mathrm{d}H.
\label{eq:conj}
\end{align}
\end{widetext}
The factor of 2 outside the first bracket in (\ref{eq:conj}) arises because a displacement of $\mathbf{r}_k^+$ induces an equivalent displacement of $\mathbf{r}_k^-$.  

\subsection{A constitutive model}

We now implement a specific mechanical energy $U(\mathsf{V}, \mathsf{a}, \mathsf{L}^+, \mathsf{L}^-, \mathsf{A}^l)$ of the monolayer.  Introducing a target volume $V_0^*$, target total area $a_0^*$ and target apical perimeter $L_0^*$, and scaling all lengths on $V_0^{*1/3}$, so that $a_0=a_0^*/V_0^{*2/3}$, $L_0=L_0^*/V_0^{*1/3}$, we write the energy (\ref{eq:genenergy}) in nondimensional form as 
\begin{multline}
U={\textstyle\sum_h} \big[\tfrac{1}{2} (V_h-1)^2 + \tfrac{1}{2}{\Gamma_a} (a_h-a_0)^2  + \tfrac{1}{2}{\Gamma_L} \big\{(L_h^+-L_0)^2 \\ +(L_h^- -L_0)^2\big\} +\tfrac{1}{2}\Gamma_A \textstyle{\sum_i} \vert C_{hi}^l\vert  A_i^l \big].
\label{eq:TotalEnery}
\end{multline} 
$\Gamma_a$, $\Gamma_L$ are positive coefficients measuring energy contributions relative to the volumetric terms.  The factor $\vert C_{hi}^l\vert$ is drawn from a block of the face-cell incidence matrix (Appendix~\ref{app:matrices}), and is unity for lateral faces at the periphery of the monolayer and 2 for all internal lateral faces; adhesion is therefore promoted by assuming the constant $\Gamma_A<0$. Here we follow convention by writing $U$ as a sum of functions that are linear or quadratic in their argument; stronger nonlinearities can be used to prevent lengths or areas from falling to zero under a finite energy change \cite{cowley2024}. The partial derivatives (\ref{eq:pderivs}) of $U$ can then be written
\begin{subequations}
\label{eq:pta}
\begin{align}
P_h&=V_h-1,& T_h&=\Gamma_a(a_h-a_0), \\
T_h^{p\pm}&=\Gamma_L(L_h^\pm -L_0), & T_i^l&=\tfrac{1}{2}\Gamma_A{\textstyle \sum_h}\vert C_{hi}^l\vert.
\end{align}
\end{subequations}
Apical-basal symmetry enforces, via (\ref{eq:geo_quants}),
$U=\sum_h \mathcal{U}(A_h^+,L_h^+,H)$ where the energy of cell $h$ is
\begin{multline}
U_h\equiv \mathcal{U}(A_h^+,L_h^+,H)=\tfrac{1}{2} (A_h^+ H-1)^2 + {\Gamma_L} (L_h^+-L_0)^2 \\ 
+ \tfrac{1}{2}{\Gamma_a} (2A_h^++HL_h^+-a_0)^2 +\tfrac{1}{2}\Gamma_A H L_h^+.
\label{eq:energy_AL}
\end{multline}
Here we have combined (\ref{eq:la}) and (\ref{eq:aa}) to give $L_h^+=\sum_i \vert C_{hi}^l \vert t_i^+$, writing cell perimeter in terms of component apical edge lengths.  Every cell now has an equivalent energy $U_h$ dependent on its apical area and perimeter, drawing a connection to traditional 2D cell vertex models (reviewed in Appendix~\ref{2D Appendix}). The pressure and tensions (\ref{eq:pta}) can be written 
\begin{subequations}
\begin{align}
P_h&=A_h^+ H-1, \quad T_h = \Gamma_a(2A_h^+ + L_h^+ H - a_0), \\
T_h^c &\equiv T_h^{p+}+T_h^{p-}=2\Gamma_L (L_h^+-L_0),
\label{eq:PTs}
\end{align}
\end{subequations}
where the effective cortical tension $T_h^c$ has apical and basal contributions.  The choice of adhesion energy allows us to combine (\ref{eq:pta}b)$_2$ with (\ref{eq:ids}) in (\ref{eq:conj}), and hence define the quantities conjugate to variations in apical areas and perimeters as 
\begin{subequations}
\label{eq:p2t2x}
\begin{align}
    P_h^{2D}&\equiv\frac{\partial{U}_h}{\partial A_{h}^+}=P_h H +2T_h \nonumber\\& =  H(A_h^+ H-1) + 2 \Gamma_a(2A_h^+ + L_h^+ H - a_0),
\label{eq:P2Da0} \\
    T_h^{2D}&\equiv \frac{\partial U_h}{\partial L_h^+}=T_h H+T_h^c +\tfrac{1}{2}\Gamma_A H \nonumber \\&= \left(\Gamma_a(2A_h^+ + L_h^+ H - a_0)+ \tfrac{1}{2} \Gamma_A\right)H\nonumber\\& \qquad +2\Gamma_L (L_h^+-L_0).
\label{eq:T2Da0}
\end{align}
\end{subequations}
Treating $U$ as a function of $H$ and $\mathbf{r}_1^+, \dots \mathbf{r}_{N_v}^+$, we can then write the in-plane and out-of-plane forces at the apical surface as, respectively, for $k=1,\dots,N_v$,
\begin{subequations}
\label{eq:p2t2}
\begin{align}
\frac{\partial U}{\partial \mathbf{r}_k^+}&=2\left(\sum_h \left\{ P^{2D}_h \frac{\partial A_h^+}{\partial \mathbf{r}_k^+} + T^{2D}_h \frac{\partial L_h^+}{\partial \mathbf{r}_k^+}\right\}\right),\\
\frac{\partial U}{\partial H}&=\sum_h\left(P_hA_h^++T_hL_h^++\tfrac{1}{2}\Gamma_A L_h^+\right).
\end{align}
\end{subequations}
For a given set of vertex locations, the global vertical force balance $\partial U/\partial H=0$ identifies the equilibrium height as
\begin{equation}
H^*= \frac{ {\sum_h} \left\{  A_h^+ +\left( \Gamma_a (a_0-2 A_h^+)-\tfrac{1}{2}\Gamma_A\right)L_h^+\right\}}{\sum_h \left\{ (A_h^+)^2+\Gamma_a (L_h^+)^2\right\} }.
\label{eq:Ha0}
\end{equation}
A more nuanced treatment of the vertical force balance is given in Appendix~\ref{sec:height}; we will use this later to evaluate patterns of weak cell height variation across a monolayer.

Finally, 
we scale $a_0$ from the equilibrium {\color{black} mechanical} problem by setting
    \begin{equation}
    \label{eq:scalea0}
\begin{gathered}
        A_h^+=a_0\tilde{A}_h,~ H=\frac{\tilde{H}}{a_0}, ~ L_h^+=a_0^2 \tilde{L}_h, ~ {\color{black}P_h^{2D}}=\frac{{\color{black}\tilde{P}_h^{2D}}}{a_0},\\  {\color{black}T_h^{2D}}=\frac{{\color{black}\tilde{T}_h^{2D}}}{a_0^2}, \,
        \Gamma_a=\frac{\tilde{\Gamma}_a}{a_0^2}, \, \Gamma_A=\frac{\tilde{\Gamma}_A}{a_0}, \, \Gamma_L=\frac{\tilde{\Gamma}_L}{a_0^4}, \, L_0=a_0^2\tilde{L}_0.
\end{gathered}
\end{equation}
{\color{black}$a_0$ is retained in the geometric ratio $L_h^+/\sqrt{A_h^+}=a_0^{3/2} \tilde{L}_h/\sqrt{\tilde{A}_h}$, which takes the value $2^{3/2}3^{1/4}$ for perfect hexagons.}  Dropping tildes, (\ref{eq:scalea0}) allows us to write (\ref{eq:p2t2x}) and \eqref{eq:Ha0} as
\begin{subequations}
\label{eq:H}
\begin{align}
P^{2D}_h&\equiv P^{2D}(A_h,L_h, H) \nonumber \\& 
= H\left(A_h H -1\right)  + 2\Gamma_a\left(2A_h+L_h H -1\right),
\label{P_2D}\\
T^{2D}_h&\equiv T^{2D}(A_h,L_h,H) =  \Gamma_a\left(2A_h+L_h H -1\right) H \nonumber \\
& \qquad\qquad \qquad \quad +\tfrac{1}{2}\Gamma_A H + 2\Gamma_L\left(L_h-L_0\right),
\label{T_2D}\\
H^* &= \frac{\sum_h \left\{ A_h +\Gamma_a\left(1-2A_h\right)L_h - \tfrac{1}{2}\Gamma_AL_h \right\} }{\sum_h \left\{ A_h^{2}+\Gamma_a L_h^{2}\right\} }.
\end{align}
\end{subequations}
These yield pressures $P_h^{2D}(A_h,L_h, H^*)$ and tensions $T_h^{2D}(A_h,L_h, H^*)$ defined exclusively in terms of apical areas $A_h$ and perimeters $L_h$ when $H$ takes its equilibrium value $H^*$.

To summarise, in typical 2D formulations of the vertex model, the cell energy is defined by $A_h$ and $L_h$ alone, giving rise to a surface tension (apical surface pressure) that is dependent on $A_h$ alone, and a cortical line tension that is dependent on $L_h$ alone (Appendix~\ref{2D Appendix}).  Here, their 3D analogues $P_h^{2D}(A_h, L_h, H)$ and $T_h^{2D}(A_h,L_h,H)$ have a more intricate functional dependence (\ref{eq:H}) because of the incorporation of bulk effects relating to changes in cell volume and total surface area.  Unlike the classical 2D pressure $A_h-1$, here the composite apical surface pressure $P_h^{2D}$ incorporates height changes and is coupled to apical perimeter variations through the total area.  Unlike the classical cortical tension $\Gamma_L(L_h-L_0)$, the composite cortical tension $T^{2D}_h$ is coupled to apical area variations.  Lateral adhesion introduces a term $\tfrac{1}{2}\Gamma_A H$ to $T_h^{2D}$ that can be considered to contribute to the target perimeter, albeit in an intricate manner.

\subsection{Monolayer configurations}
\label{sec:config}

To understand the properties of the model, it is helpful to investigate some specific monolayer configurations, which we will illustrate numerically in Sec.~\ref{sec:results} below.

For an array of identical hexagonal prismatic cells of edge-length $l$, $A_h\equiv A=\tfrac{1}{2}3\sqrt{3} {\color{black}a_0^3}l^2$, $L_h\equiv L=6l$ and the horizontal force (\ref{eq:p2t2}a) becomes 
\begin{equation}
a_0\frac{\partial U}{\partial \mathbf{r}_k^+}={\color{black}2}\sum_h \left\{ 3\sqrt{3}{\color{black}a_0^3}l P^{2D}_h  + 6 T^{2D}_h \right\}\frac{\partial l}{\partial \mathbf{r}_k^+}.
\label{eq:p2t2a}
\end{equation}
The forces due to apical area and apical perimeter changes are parallel in (\ref{eq:p2t2a}), leading to the coupled algebraic system for equilibrium configurations:
\begin{subequations}
\label{eq:hxf}
\begin{align}
    0&=2A P^{2D}(A,L,H)  + L T^{2D}(A,L,H), \\  A&=\tfrac{1}{2}3\sqrt{3}{\color{black}a_0^3}l^2, \quad L=6l,\\
    H &= \frac{ A +\Gamma_a\left(1-2A\right)L - \tfrac{1}{2}\Gamma_A L  }{  A^{2}+\Gamma_a L^{2} }.
\end{align}
\end{subequations}
Elimination of $H$, $A$ and $T$ from (\ref{eq:H}a,b) and (\ref{eq:hxf}) yields a single nonlinear scalar equation, with roots specifying possible values of $l$.  Some asymptotic limits of configurations for rigid hexagons are described in Appendix~\ref{app:limits}. As in the 2D model (Appendix~\ref{2D Appendix}; Fig.~\ref{fig:2D_fig} below), physical solutions are limited to certain regions of parameter space.  We restrict attention to solutions with $l>0$ and $H>0$; where multiple equilibria exist, we consider only solutions that are stable when vertices move under a frictional drag, as determined by analysis of eigenvalues of the relevant Hessian (Appendix~\ref{app:dynamics}).  We expect solutions for which $T^{2D}>0$ to be rigid.  

We can alternatively seek solutions for which cells need not be identical, but for which the system is floppy in the plane, while still satisfying the vertical force balance.  In this instance, the problem is determined by 
\begin{subequations}
\label{eq:flo}
\begin{align}
P^{2D}(A,L,H)=0, \quad T^{2D}(A,L,H)&=0, \\  (AH-1)A+[\Gamma_a(2A+LH-1)+\tfrac{1}{2}\Gamma_A] L&=0.   
\end{align}
\end{subequations}
Solutions for $n$-sided cells will satisfy the isoperimetric inequality 
\begin{equation}
    \frac{{\color{black}a_0^{3/2}} L}{\sqrt{A}}\geq s_n=2\sqrt{n \tan\left(\frac{\pi}{n}\right)},
    \label{eq:isoperim}
\end{equation}
with equality arising when cells are symmetric $n$-gons.  Thus for a monolayer of hexagons, the transition between rigid and floppy states takes place when  
\begin{equation}
{\color{black}a_0^{3/2}}L/\sqrt{A}=s_6\approx 3.72,
\label{eq:s6}
\end{equation}
which is equivalent to (\ref{eq:hxf}b).  

In addition to reporting numerical solutions of (\ref{eq:flo}), and where appropriate of (\ref{eq:s6}), plus relevant asymptotic limits (Appendix~\ref{app:limits}), we also report distributions of the shape parameter $L_h/\sqrt{A_h}$ for isolated disordered monolayers, to reveal the nature of the rigidity transition.  To determine these, we used the \texttt{VertexModel.jl} package~\cite{Revell_VertexModel_jl_2022} to simulate 500-cell monolayers. {\color{black}Isolated} monolayers with stress-free external boundary conditions were grown with stochastic cell divisions, {\color{black} taking place uniformly randomly across the monolayer}, allowing T1 transitions to occur when edge lengths fell below a threshold of 0.01 \cite{kursawe2017impact}, resulting in a disordered system.  {\color{black}Once $N_c=500$, we suppressed divisions and the monolayer was relaxed to the nearest equilibrium.  In these simulations,} we replaced the tension and pressure in the 2D vertex model (\ref{eq:H}a,b) with $T^{2D}$ and $P^{2D}$, enforcing the vertical force balance (\ref{eq:H}c). Monolayers were relaxed to equilibrium with a tolerance (in vertex location) of $5\times10^{-6}$.  Simulations are reported in Section~\ref{sec:stress} below.

\subsection{Cell-level stress}
\label{sec:stress}

We derive the stress $\boldsymbol{\sigma}_h$ in a cell within the monolayer in Appendix~\ref{app:stress}, finding that
\begin{subequations}
    \label{eq:Vhsigma_3Da}
\begin{align}
V_h \boldsymbol{\sigma}_h &= A_h \boldsymbol{\sigma}_h^{2D}
+\frac{\partial U_h}{\partial H} H \hat{\mathbf{z}}\otimes \hat{\mathbf{z}},\\ \mathrm{where}\quad A_h \boldsymbol{\sigma}_h^{2D}&\equiv P_h^{2D} A_h\mathsf{I}_\perp + T_h^{2D}L_h\mathsf{Q_{\perp}}_h.
\end{align}
\end{subequations}
Here $\mathsf{I}_\perp\equiv\mathrm{diag}(1,1,0)$ and $\mathsf{Q}_\perp$ is a shape tensor satisfying $L_h\mathsf{Q_{\perp}}_h =\sum_j D^{\mathrm{ce}+}_{hj}l_j^+ \hat{\mathbf{t}}_{j}^+\otimes \hat{\mathbf{t}}_{j}^+$, so that $\mathrm{Tr}(\mathsf{Q}_{\perp h})=1$. $\mathsf{D}^{\mathrm{ce}+}$ is a block of the cell-edge adjacency matrix (see Appendix~\ref{app:matrices}); for a derivation of the in-plane stress, $\boldsymbol{\sigma}^{2D}_h$, see \cite{nestor2018}.  The in-plane isotropic and shear-stress components of $\boldsymbol{\sigma}^{2D}_h$ are, respectively \cite{nestor2018, jensen2020},
\begin{subequations}
\label{eq;peff2d}
\begin{align}
    P_h^{\mathrm{eff},2D}&\equiv \tfrac{1}{2}\mathrm{Tr}(\boldsymbol{\sigma}^{2D})=P_h^{2D}+\frac{T_h^{2D}L_h}{2A_h}, \\
\zeta_h^{2D}&\equiv  \frac{T^{2D}_h L_h}{A_h}\sqrt{-\mathrm{det}\left(\mathsf{Q_\perp}_h-\tfrac{1}{2}\mathsf{I}_\perp\right)}.
\end{align}
\end{subequations}
$\zeta_{2D}$ is the magnitude of the deviatoric component of $\boldsymbol{\sigma}_h^{2D}$.  $\mathsf{Q}_{\perp h}-\tfrac{1}{2}\mathsf{I}_\perp$ in (\ref{eq;peff2d}b) measures anisotropy in the shape of the apical face, and resistance to in-plane shear requires $T_h^{2D}>0$.  Recall that $\partial U_h/\partial H=0$ when the cell is in equilibrium.  This condition holds for identical hexagons at equilibrium, but will not hold in general for a disordered monolayer because $H$ is determined by a global rather than a local force balance; thus we retain this term in (\ref{eq:Vhsigma_3Da}a).  As (\ref{eq:Vhsigma_3Da}a) illustrates, stress components in 2D differ dimensionally from those in 3D by a lengthscale: when referring to 2D components, we are implicitly considering them as stress resultants, {\hbox{i.e.}} stress integrated over the depth of the cell.

$\boldsymbol{\sigma}_h$ in (\ref{eq:Vhsigma_3Da}) can be split into isotropic and deviatoric components by writing $\boldsymbol{\sigma}_h=P^{\mathrm{eff}}_h\mathsf{I}_3+\boldsymbol{\sigma}_h^d$ where $\mathsf{Tr}(\boldsymbol{\sigma}_h^d)\equiv 0$.  
Taking the trace of (\ref{eq:Vhsigma_3Da}a), it follows that the 3D isotropic stress is 
\begin{subequations}
    \label{eq: Peff}    
\begin{align}
    P^{\mathrm{eff}}_h&= \frac{1}{3 V_h}\left( 2 P^{2D}_h A_h + T^{2D}_h L_h + \frac{\partial U_h}{\partial H} H\right)\\
    & =\frac{2}{3H}P^{\mathrm{eff},2D}+\frac{1}{3A_h}\frac{\partial U_h}{\partial H}.
\end{align}
\end{subequations}
Thus only $2/3$ of $P^{\mathrm{eff},2D}_h/H$ contributes to $P^{\mathrm{eff}}_h$.  The bulk deviatoric stress is 
\begin{multline}
        V_h \boldsymbol{\sigma}^d_h= \tfrac{1}{3} \left(P^{2D}_h A_h -T^{2D}_h L_h - \frac{\partial U_h}{\partial H} H\right) \mathsf{I}_\perp \\+ T^{2D}_h L_h \mathsf{Q_\perp}_h +\tfrac{1}{3} \left( 2 \frac{\partial U_h}{\partial H}  H - 2 P^{2D}_h A_h -T^{2D}_h L_h\right) \hat{\mathbf{z}}\otimes \hat{\mathbf{z}}.
        \label{eq: sigmq dev}
\end{multline}
This shows how a cell under external compression ($\partial U_h/\partial H< 0$) experiences a bulk shear stress proportional to $\mathsf{S}_b\equiv \mathrm{diag}(-1,-1,2)=2\hat{\mathbf{z}}\otimes \hat{\mathbf{z}}-\mathsf{I}_\perp$, combining vertical compression with horizontal expansion.

Because $\boldsymbol{\sigma}_h$ in (\ref{eq:Vhsigma_3Da}a) has block matrix form, the characteristic equation for its eigenvalues $\lambda$ can be written as
\begin{align}
    \mathrm{det}\left(\frac{1}{H}\boldsymbol{\sigma}_h^{2D}-\lambda \mathsf{I}_\perp\right)\left( \frac{1}{A_h} \frac{\partial U_h}{\partial H} -\lambda\right)=0.
\end{align}
Thus the eigenvalues of $\boldsymbol{\sigma}_h$ are
\begin{align}
\label{eq:evs}
\lambda_0=\frac{1}{A_h} \frac{\partial U_h}{\partial H} \quad \mathrm{and}\quad
\lambda_\pm= \frac{1}{H}P^{\mathrm{eff}, 2D}_h \pm\zeta_h,
\end{align}
where $\zeta_h={\zeta_h^{2D}}/{H}$.  We can therefore diagonalise $\boldsymbol{\sigma}_h$, in the basis provided by its eigenvectors, writing it as
\begin{subequations}
\label{eq:stresscomponents}
\begin{align}
\boldsymbol{\sigma}_h    &= \tfrac{1}{3}(\lambda_0+\lambda_++\lambda_-)\mathsf{I}_3
    + \tfrac{1}{2}(\lambda_+-\lambda_-) \mathsf{S}_a \nonumber\\
& \qquad
+ \tfrac{1}{6}(2\lambda_0-(\lambda_+ +\lambda_-))
 \mathsf{S}_b
  \\
    &=P_h^\mathrm{eff} \mathsf{I}_3
    +\zeta_h \mathsf{S}_a
    + \left(P^{\mathrm{eff}}_h-(P^{\mathrm{eff}, 2D}_h/H)\right)\mathsf{S}_b,
\end{align}
\end{subequations} 
where $\mathsf{S}_a\equiv \mathrm{diag}(1,-1,0)$. $\zeta_h$ (see (\ref{eq:evs})) gives the magnitude of the in-plane shear stress in terms of the shear-stress resultant $\zeta_h^{2D}$ \cite{jensen2020} defined using the apical surface alone. The remaining shear-stress magnitude, $P^{\mathrm{eff}}_h-P^{\mathrm{eff}, 2D}_h/H$, characterises the difference between 3D dilation and in-plane dilation. The 2D isotropic stress $P^{\mathrm{eff},2D}_h$ and 2D shear stress $\zeta_h^{2D}$ determine the bulk stress via 
\begin{equation}
\label{eq:strf}
H \boldsymbol{\sigma}_h 
    =\tfrac{2}{3}P_h^{\mathrm{eff},2D} \mathsf{I}_3
    + \zeta_h^{2D} \mathsf{S}_a
    - \tfrac{1}{3} P^{\mathrm{eff},2D}_h \mathsf{S}_b
    +\frac{H}{A}\frac{\partial U_h}{\partial H} \hat{\mathbf{z}}\otimes\hat{\mathbf{z}},
\end{equation}
with respect to coordinates aligned with the principal axes of shape of the apical face \cite{nestor2018} and $\hat{\mathbf{z}}$.  A cell with $P_h^{\mathrm{eff},2D}<0$ is compressed (by its surroundings or membrane tension); this generates a bulk shear stress having an expansive component normal to the plane.

\subsection{Tissue-level stress and external loading}

Having determined the stress of individual cells, it is straightforward to integrate to tissue level, and thereby to consider the response of cells to an external load.  Consider a cluster $G$ of connected cells identified by the chain $\mathsf{g}=(g_1,\dots,g_{N_c})$ with elements $g_h=1$ for $h$ labelling cells in $G$ and $g_h=0$ otherwise.  These cells have total apical area $A_G=\sum_h g_h A_H$ and volume $V_G=\sum_h g_h V_h$.  Using (\ref{eq:Vhsigma_3Da}a), the total stress of the cluster, $\boldsymbol{\sigma}_G$, satisfies \cite{nestor2018}
\begin{equation}
V_G\boldsymbol{\sigma}_G\equiv {\textstyle{\sum_h}}g_h V_h{\boldsymbol{\sigma}_h}={\textstyle{\sum_h}}g_h\left(A_h {\boldsymbol{\sigma}_h^{2D}}+\frac{\partial U_h}{\partial H} H \hat{\mathbf{z}}\otimes\hat{\mathbf{z}}\right).
\end{equation}
Imposing an external load on lateral (but not apical or basal) faces using ${\boldsymbol{\sigma}_G}=P_{\mathrm{ext}}\mathsf{I}_\perp$, to mimic cell crowding in a confined environment, it follows that 
\begin{equation}
\label{eq:clus}
A_G H P_{\mathrm{ext}}\mathsf{I}_\perp = {\textstyle{\sum_h}}g_h A_h {\boldsymbol{\sigma}_h^{2D}}.
\end{equation}
The vertical force $\sum_h g_h \partial U_h/\partial H$ will vanish when $\mathsf{g}=\mathsf{1}\equiv (1,1,\dots,1)$, {\hbox{i.e.}} when $G$ is the whole monolayer, or for more general clusters taken from a large monolayer of near-identical hexagons.  In the latter case, (\ref{eq:clus}) simplifies to $P^{\mathrm{eff},2D}_h\approx HP_{\mathrm{ext}}$ (ignoring variations at the monolayer periphery).  To capture the response of cells to in-plane compression, we replace (\ref{eq:hxf}a) with $P^{\mathrm{eff},2D}=HP_{\mathrm{ext}}$ and use (\ref{eq:H}a,b), (\ref{eq:hxf}b, \ref{eq:hxf}c) and (\ref{eq;peff2d}) to solve for $A$, $L$, and $H$.  Using (\ref{eq:energy_AL}), this is equivalent to the algebraic system
\begin{equation}
    2A\frac{\partial \mathcal{U}}{\partial A}+L\frac{\partial \mathcal{U}}{\partial L}=2AHP_{\mathrm{ext}}, ~~ \frac{\partial \mathcal{U}}{\partial H}=0, ~~ A=\frac{\sqrt{3}}{24}{\color{black}a_0^3}L^2.
    \label{eq:lateralcomp}
\end{equation}
In contrast, isotropic external compression of a monolayer of near-identical hexagons via ${\boldsymbol{\sigma}_G}=P_{\mathrm{ext}}\mathsf{I}_3$ leads to the constraints $P^{\mathrm{eff},2D}=HP_{\mathrm{ext}}$ and $\partial \mathcal{U}/\partial H=A P_{\mathrm{ext}}$, requiring modification of (\ref{eq:hxf}a) and (\ref{eq:hxf}c).  This is equivalent to the system {\color{black}(\ref{eq:lateralcomp}) but with $\partial U/\partial H=0$ replaced by}
\begin{equation}
    \frac{\partial \mathcal{U}}{\partial H}=A P_{\mathrm{ext}}.
    \label{eq:isotropiccomp}
\end{equation}
We report numerical solutions of (\ref{eq:lateralcomp}) and (\ref{eq:isotropiccomp}) below, to demonstrate the nonlinear response of cells to simple loading.  Responses of cells to an arbitrary small-amplitude strains, captured as prestress-dependent stiffnesses, are discussed in Appendix~\ref{app:stiffness}. 

\section{Results}
\label{sec:results}

We first discuss properties of an array of identical hexagonal cells, examining the impact of variation of individual parameters and of external loading, before turning to a disordered monolayer.  

\subsection{Regular hexagonal cells: parameter variation} 

A periodic array of regular hexagonal cells, in the rigid phase, will have uniform tension and pressure at equilibrium, making the solution for a single cell sufficient to describe the whole array.  We use this simple case to illustrate the general behaviour of the model.  We solve (\ref{eq:hxf}) with (\ref{eq:H}a,b) for cells in the rigid phase, disregarding solutions for which $A\leq 0$, $L\leq 0$ or $H\leq 0$. When cells lose in-plane rigidity, equilibrium solutions (which will no longer be {regular hexagons}) are found from \eqref{eq:flo}.  As solutions satisfy nonlinear algebraic equations, we use root-tracing to identify distinct solution branches; their stability is verified using the Hessian given in Appendix~\ref{app:dynamics}.

\begin{table}
\begin{ruledtabular}
    \begin{tabular}{lcc|lcc}
     {Quantity}  && {Value}   & {Quantity}  && {Value}\\
    \hline
    Volume& $V_h$   & 0.49  & Bulk pressure& $P_h$     & -0.51 \\    
    Total surface area& $a_h$ &4.18  & Surface tension& $T_h$  &0.32 \\
       Apical area& $A_h$    &  0.24   &  Cortical tension& $T^c_h$     & -0.03\\
    Apical perimeter& $L_h$   & 1.84    &  \\
     Monolayer height& $H$    & 2.01 & 2D pressure& $P^{2D}_h$     & -0.39\\
    Edge length & $l$     & 0.31 & 2D tension& $T^{2D}_h$     &0.10\\
    \end{tabular}
    \caption{Summary of approximate values of geometric and mechanical quantities for hexagonal cells in the baseline case, with $\Gamma_a=\Gamma_L=0.1, \Gamma_A=-0.5$, $L_0=2$ {\color{black}and $a_0=1$}.}
    \label{tab:base_case_quatities}
    \end{ruledtabular}
\end{table}

We choose a baseline case where $\Gamma_a=\Gamma_L=0.1, \Gamma_A=-0.5$, $L_0=2$ {\color{black}and $a_0=1$}, yielding geometric and mechanical variables listed in Table~\ref{tab:base_case_quatities}.  This example is in the rigid regime, with positive tension, $T^{2D}_h$, balancing negative pressure, $P^{2D}_h$; negative cortical tension $T_h^c$ is offset by positive surface tension $T_h$. We found only one solution for this set of parameters. We examine variations of each parameter in turn before taking a wider view of parameter space.  In the 2D quadratic model \cite{farhadifar2007}, hexagons lose rigidity when the target apical perimeter exceeds $s_6\approx 3.72$ (Appendix~\ref{2D Appendix}).  In the present 3D model, other factors contribute to $T_h^{2D}$, and thus to the rigidity transition.

\begin{figure}
    \centering
    \includegraphics[width=\linewidth]{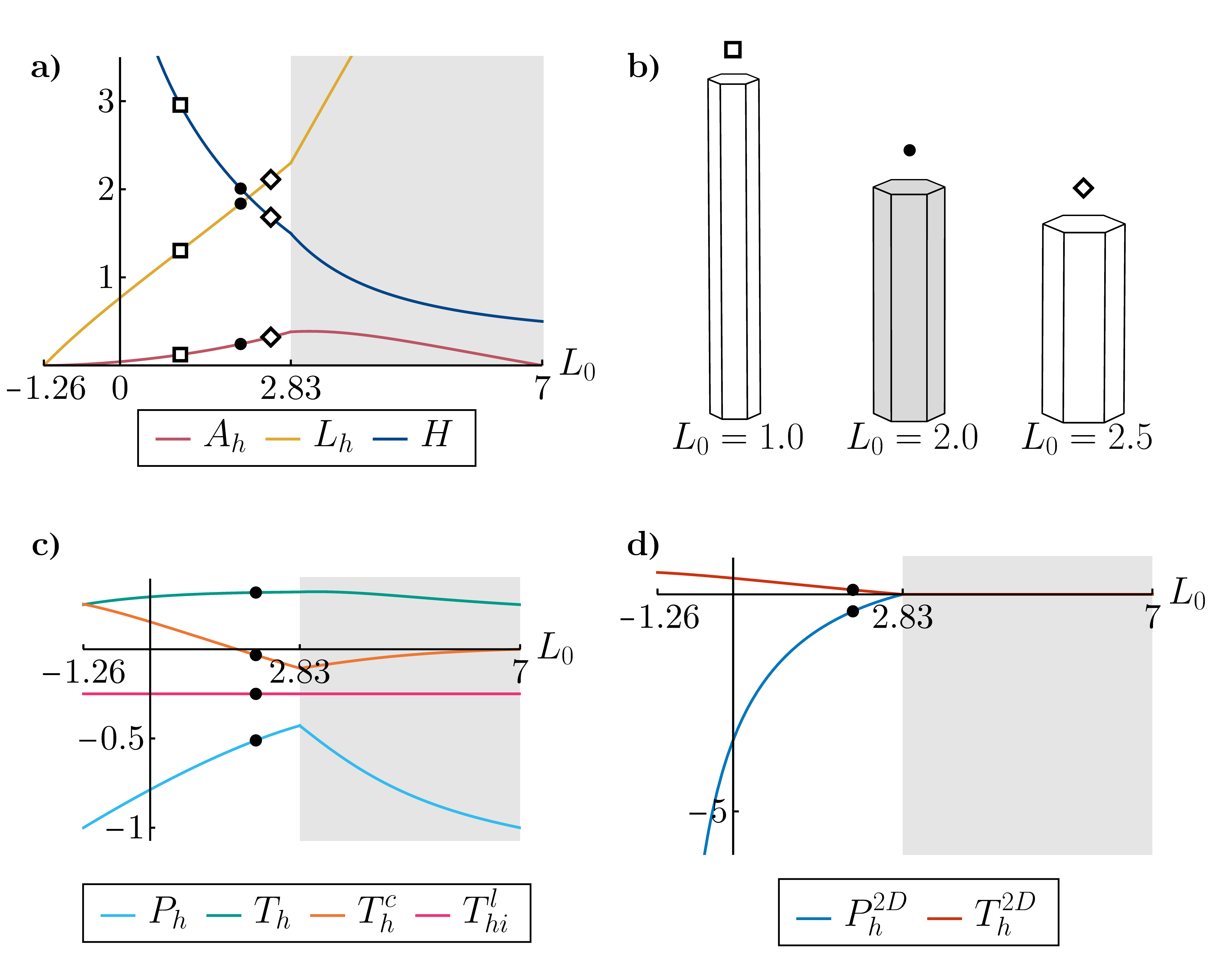}
        \caption{Impact of variation of $L_0$.  (a) Cell apical area, $A_h$, perimeter, $L_h$, and height, $H$ as functions of $L_0$. Symbols correspond to the cell shapes illustrated in (b). The baseline cell shape is shown in grey (solid symbol); cells for $L_0={1,2.5}$ are marked by open symbols. (c) Bulk pressure, $P_h$ (cyan), bulk tension, $T_h$ (green), cortical line tension, $T_h^c$ (orange), contribution of adhesive tension from a cell's lateral face, $T_{hi}^l=\frac{1}{2} \Gamma_A\vert C_{hi}^l\vert$ (pink), and (d) $P^{2D}_h$ (blue), $T^{2D}_h$ ({\color{black}brown}) versus $L_0$. In all panels, black points at $L_0=2$ indicate the baseline case, the grey shaded region indicates the floppy regime, the hatched regions indicate no physical solutions, and $\Gamma_a=\Gamma_L=0.1, \Gamma_A=-0.5$, {\color{black}$a_0=1$}. }
        \label{fig:Vary_L0}
\end{figure}

Figure \ref{fig:Vary_L0} shows how the cell shape, pressures and tensions change with $L_0$, with other parameters held at their baseline values.  In the rigid regime, increasing $L_0$ leads, intuitively, to an increase in apical area and perimeter and a decrease in height (Fig.~\ref{fig:Vary_L0}a), making cells more squamous. As shown in Fig.~\ref{fig:Vary_L0}(d), $T^{2D}$ and $P^{2D}$ fall to zero at $L_0\approx 2.83$, resulting in a loss of in-plane rigidity. Beyond the rigidity transition, the apical area (of distorted floppy cells) decreases, going to zero at $L_0\approx 7.0$; the perimeter and height continue to increase and decrease respectively.  In contrast, reducing $L_0$ constricts the apical face until, at $L_0\approx-1.26$, the equilibrium area and perimeter vanish.  Thus rigid and floppy solutions exist only over a finite range of $L_0$ values, and highly elongated (columnar) cells are readily obtained by a reduction in $L_0$.

\begin{figure}
     \centering
         \includegraphics[width=\linewidth]{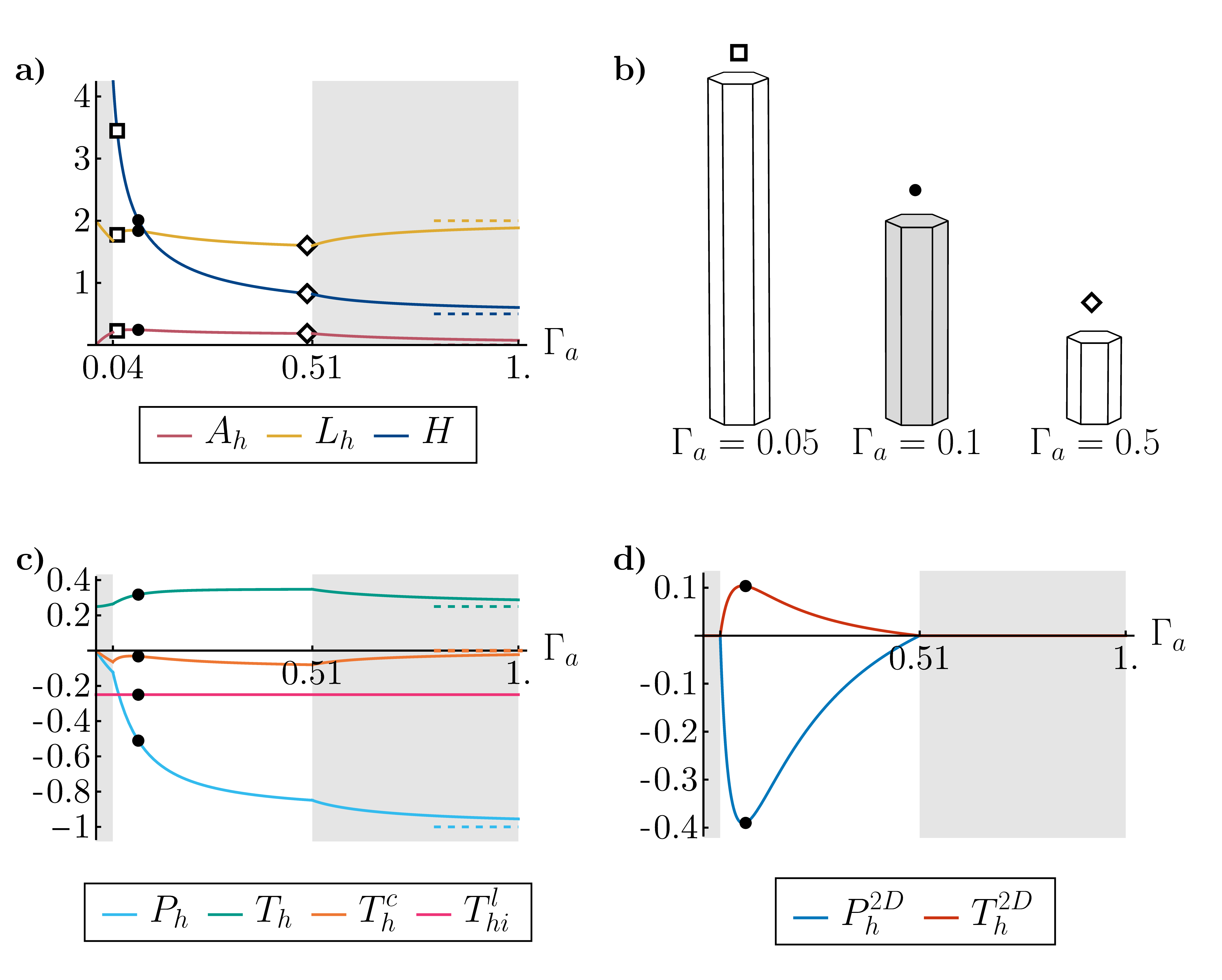}
        \caption{Impact of variation of $\Gamma_a$, presented following the format of Fig.~\ref{fig:Vary_L0}. (b) The baseline cell shape at $\Gamma_a=0.1$ is shown in grey (solid symbol); cells for $\Gamma_a={0.05,0.5}$ are marked by open symbols. In all panels, $\Gamma_L=0.1, \Gamma_A=-0.5, L_0=2.0$ {\color{black}and $a_0=1$}. The dashed lines in (a,c) show the asymptote {\color{black}(\ref{eq:elong})} as $\Gamma_a \rightarrow \infty$. 
        }
        \label{fig:Vary_Gamma_a}
\end{figure}

$\Gamma_a$ controls the strength of the bulk tension associated with variations of total surface area. This term contributes to the compound pressure and tension terms at the apical surface, and is responsible for coupling $A_h$, $L_h$ and $H$ in determining $P^{2D}$ and $T^{2D}$.  Figure \ref{fig:Vary_Gamma_a} shows the impact of  changing $\Gamma_a$. In the rigid regime, increasing $\Gamma_a$ causes the cell to become more cuboidal, consistent with expectation of increasing the overall surface tension.  In-plane rigidity can be lost either by decreasing or, less intuitively, increasing $\Gamma_a$ past thresholds.  As $\Gamma_a$ increases, the more squamous floppy solution asymptotes quickly towards the limiting case of fixed surface area (\ref{eq:larga}, \ref{eq:elong}), achieved by strong in-plane distortions.  When $\Gamma_a \rightarrow 0$ the cell area in the more columnar floppy state shrinks to zero, $P_h$ and $T_h^c$ go to zero and the bulk and adhesive tensions, $T_h^c$ and $T_{hi}=\tfrac{1}{2}\Gamma_A\vert C_{hi}^l\vert$, balance each other. 

\begin{figure}
     \centering
         \includegraphics[width=\linewidth]{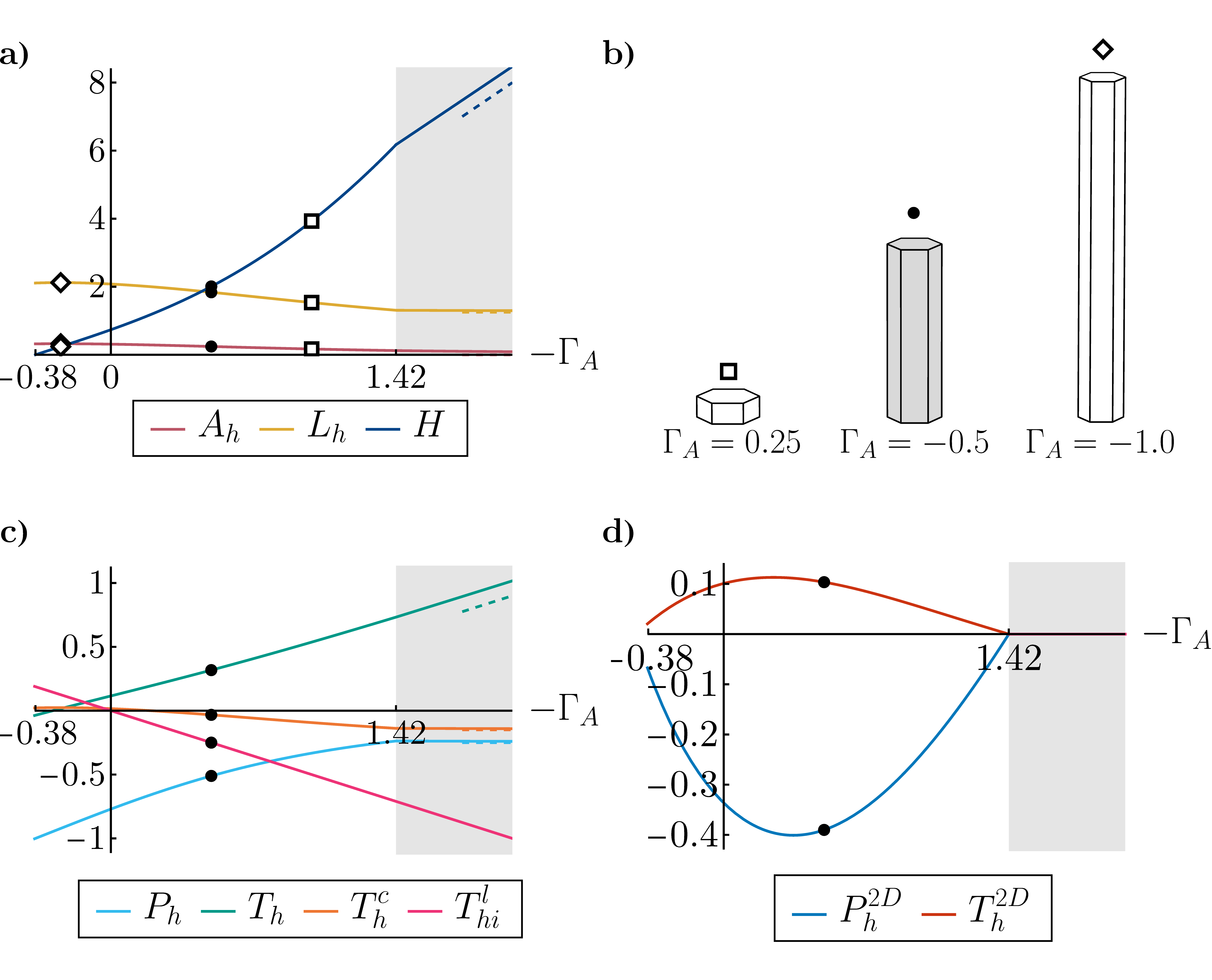}
        \caption{Impact of varying $-\Gamma_A$, presented following the format of Fig.~\ref{fig:Vary_L0}. (b) The baseline cell shape at $\Gamma_A=-0.5$ is shown in grey (solid symbol); cells for $\Gamma_A={0.25,-1.0}$ are marked by open symbols. In all panels, $\Gamma_a=\Gamma_L=0.1, L_0=2.0$ {\color{black}and $a_0=1$}.   The dashed lines in (a,c) show the asymptote (\ref{eq:gaf}) as $-\Gamma_A \rightarrow \infty$.}
        \label{fig:Vary_Gamma_A}
\end{figure}

$-\Gamma_A$ controls the strength of cell-cell adhesion, so that increasing $-\Gamma_A$ acts to increase lateral face area.  Accordingly, decreasing $-\Gamma_A$ penalises contact to the extent that $H$ falls to zero at $-\Gamma_A\approx -0.38$ (Figure \ref{fig:Vary_Gamma_A}). While the cell height monotonically increases with $-\Gamma_A$, there is a maximum in apical area and perimeter at $-\Gamma_A\approx -0.25$.  Strong adhesion elongates cells but also drives them into a floppy regime.  This can be understood from the terms $\tfrac{1}{2}\Gamma_A H -2 \Gamma_L L_0$ in $T^{2D}$ in (\ref{eq:H}c): increasing $-\Gamma_A$ has an effect roughly equivalent to increasing $L_0$, which as already been seen to induce loss of rigidity.  Once in the floppy regime, the asymptote (\ref{eq:gaf}) is recovered, with $H\propto -\Gamma_A$, $A\propto -\Gamma_A^{-1}$ and $L\propto -\Gamma_A^0$, consistent with very strong adhesion driving the formation of columnar cells.  We will show below how columnar cells can also remain rigid for large $-\Gamma_A$.

To confirm the nature of the rigidity transitions in Figs~\ref{fig:Vary_L0}--\ref{fig:Vary_Gamma_A}, we tracked eigenvalues of the Hessian (\ref{eq:hess}) while varying $L_0$, $\Gamma_a$ and $\Gamma_A$ across rigidity boundaries (Appendix~\ref{app:dynamics}; Fig.~\ref{fig:evals} below).  Assuming the cell height relaxes rapidly to equilibrium, a hexagonal cell has 12 degrees of freedom (two per vertex) and the Hessian therefore has 12 eigenvalues.  Three of these are degenerate, representing translations and rotation that do not alter the cell's energy.  The remainder involve a dilational mode and 8 symmetry-breaking eigenmodes.  At a rigidity transition, these 8 modes become unstable (Fig.~\ref{fig:evals}).  Thus the rigidity transitions shown in Figs~\ref{fig:Vary_L0}--\ref{fig:Vary_Gamma_A} mark the bifurcation of 8 floppy states from the hexagonal state, with all 8 states sharing the same values of apical area and perimeter.

\begin{figure}
     \centering
         \includegraphics[width=\linewidth]{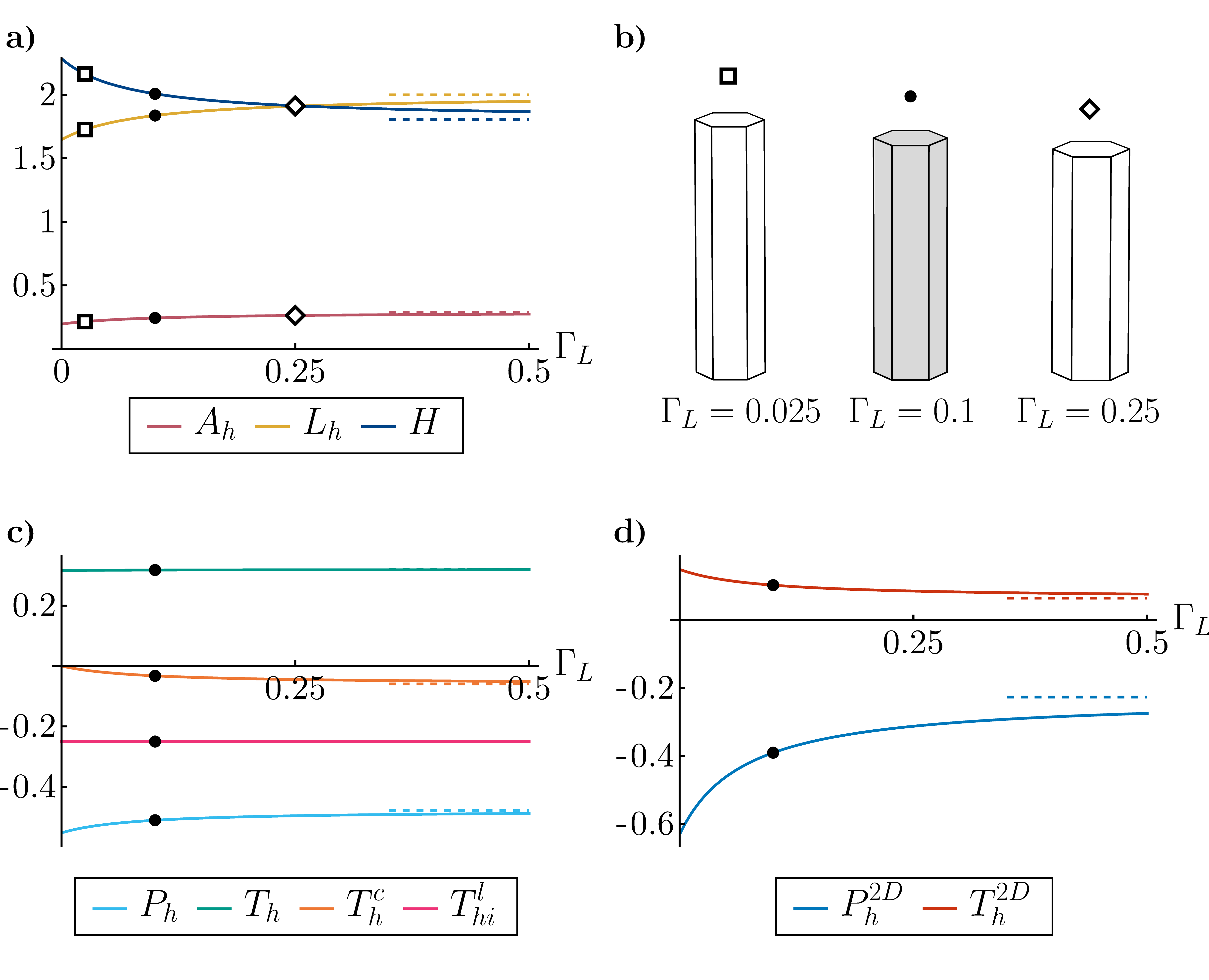}
        \caption{Impact of varying $\Gamma_L$, presented following the format of Fig.~\ref{fig:Vary_L0}. (b) The baseline cell shape at $\Gamma_L=0.1$ is shown in grey (solid symbol); cells for $\Gamma_L={0.025,0.25}$ are marked by open symbols. In all panels, $\Gamma_a=0.1, \Gamma_A=-0.5,  L_0=2.0$ {\color{black}and $a_0=1$.} The dashed lines show the asymptote as $\Gamma_L \rightarrow \infty$ as calculated in \eqref{eq:largl}.}
        \label{fig:Vary_Gamma_L}
\end{figure}

$\Gamma_L$ is the parameter that most closely resembles the apical tension parameter $\Gamma_{2D}$ in the 2D model (Appendix~\ref{2D Appendix}).  It controls the strength of the cortical line tension at the apical/basal perimeter, $T_h^c$. The term containing $\Gamma_L$ in $T_h^{2D}$ is the only term independent of $H$.  For the chosen parameters in Fig.~\ref{fig:Vary_Gamma_L}, the cell remains in the rigid regime for $\Gamma_L\geq 0$. There is a small variation in shape and mechanical quantities before they asymptote to (\ref{eq:largl}) as $\Gamma_L\rightarrow\infty$, with the apical perimeter constrained to be very close to $L_0$.  As we will see shortly, for other parameter choices, the cell can enter a floppy regime with increasing $\Gamma_L$.  
 
{\color{black}Figure~\ref{fig:vary_a0} (Appendix~\ref{app:tsa} below) shows the impact of varying the target surface area $a_0$ from unity with other parameters held at their baseline values.  Increasing $a_0$ takes the system into a floppy state, while causing cells to elongate.}

\begin{figure}
\centering
\includegraphics[width=\linewidth]{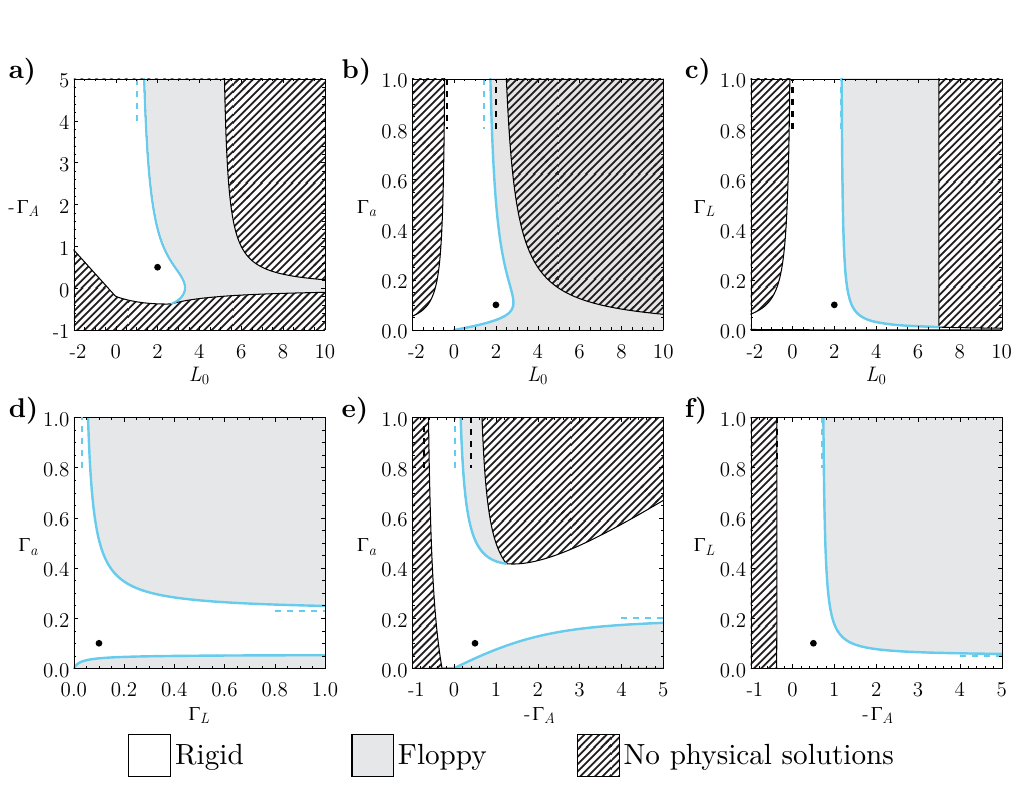}
\caption{Cross-sections of $(L_0, \Gamma_a, -\Gamma_A, \Gamma_L)$-parameter space, intersecting the baseline case (black circle).  In each plot, only two parameters are varied.  Hatching shows regions where there are no physical solutions. Grey regions show the floppy regime.  Blue lines show where $T^{2D}=0$, corresponding to a rigidity transition; blue dashed lines show asymptotes at $\Gamma_a \rightarrow \infty,\Gamma_A \rightarrow -\infty$ and $\Gamma_L \rightarrow \infty$, calculated using equations \eqref{eq:larga}, \eqref{eq:floplimlargeGa} and \eqref{eq:largl} respectively.  Black dashed lines show asymptotes marking where $H\rightarrow 0$ (for large $\Gamma_L$, using (\ref{eq:sqlim}), in (f)) or where $H\rightarrow \infty$ (for large $\Gamma_a$, using (\ref{eq:elonga}b), in (b, e); and for large $\Gamma_L$, using (\ref{eq:colim}, in (c)).}
\label{fig:rigidity transition}
\end{figure}

The transitions described above are summarized in Fig.~\ref{fig:rigidity transition}, which shows six 2D slices through parameter space, for hexagonal cells. In comparison to the standard 2D model, there are multiple paths in parameter space along which the system loses rigidity, or along which rigid solutions become unphysical.  The $(L_0,\Gamma_L)$-plane (Fig.~\ref{fig:rigidity transition}c) is the most direct analogue of the standard 2D model (Fig.~\ref{fig:2D_fig}), but illustrates significant variation in the value of $L_0$ at which the rigidity transition takes place.  Figure~\ref{fig:rigidity transition}(a) offers an example where rigid solutions are lost either because the apical area vanishes (by decreasing $L_0$ from the baseline value, for fixed $-\Gamma_A$), or because of the termination of a steady solution branch at a saddle-node bifurcation (by simultaneous reduction of $L_0$ and $-\Gamma_A$), at which the eigenvalue corresponding to the dilational mode of the Hessian vanishes. Figure~\ref{fig:rigidity transition}(e) shows how variation of both $-\Gamma_A$ and $\Gamma_a$ can lead to loss of rigidity (increasing $-\Gamma_A$ from the baseline case), flattening of the cell (reduction of $-\Gamma_A$) or cell elongation (via simultaneous increase of $-\Gamma_A$ and $\Gamma_a$).  Fig.~\ref{fig:rigidity transition}(d,e) demonstrates distinct floppy states, either squamous (low $\Gamma_a$) or columnar (high $\Gamma_a$); these are reflected in the two branches of the rigid/floppy boundary in Fig.~\ref{fig:rigidity transition}(a,b).  {\color{black}Figure~\ref{fig:a0_rigidity} (Appendix~\ref{app:tsa} below) maps out variations of $a_0$ versus $\Gamma_a$, $-\Gamma_A$, $\Gamma_L$ and $L_0$.  For large $\Gamma_a$ or large $\Gamma_L$ (constraining the total area or perimeter), rigid solutions are confined to a window of $a_0$ values, demonstrating that rigidity can be lost by a reduction in $a_0$.}

We used asymptotic approaches (Appendix~\ref{app:limits}) to identify the limiting location of the rigid/floppy transition, such as the limit $\Gamma_a\approx \Gamma_L L_0$ (\ref{eq:floplimlargeGa}) for large $-\Gamma_A$ (Fig.~\ref{fig:rigidity transition}a,e,f), (\ref{eq:larga}) for large $\Gamma_a$ (Fig.~\ref{fig:rigidity transition}b,d,e {\color{black} and \ref{fig:a0_rigidity}a}) and (\ref{eq:largl}) for large $\Gamma_L$ (Fig.~\ref{fig:rigidity transition}c,d,f {\color{black} and \ref{fig:a0_rigidity}c}).  The boundary at which rigid cells become strongly elongated in Fig.~\ref{fig:rigidity transition}(b) is $L_0\approx -\sqrt{3}/(48\Gamma_L)$ (\ref{eq:elonga}b)$_3$.  Similarly, rigid cells become strongly flattened at the threshold (\ref{eq:sqlim}) for large $\Gamma_L$ (Fig.~\ref{fig:rigidity transition}(f); the boundary turns out to be almost independent of $\Gamma_L$) and columnar at $\Gamma_a\approx -L_0^3/192$ (\ref{eq:colim}) (Fig.~\ref{fig:rigidity transition}c).  These limiting approximations reflect dominant physical balances.  Cells can remain rigid for large $-\Gamma_A$ when $\Gamma_a$ is also large (Fig.~\ref{fig:rigidity transition}e), with the associated tensions balancing to determine $H\approx \sqrt{3}\Gamma_A^2/(96\Gamma_a^2)$ (\ref{eq:gar}).  In none of these limits does the 3D model recover results from the classical 2D model. 

{\color{black}The slices of parameter space in Figs~\ref{fig:rigidity transition} and \ref{fig:a0_rigidity} do not access a further limiting case, namely that of constrained cell volume, for which $\Gamma_a$, $\vert\Gamma_A\vert$ and $\Gamma_L$ are all small in magnitude.  We show in Appendix~\ref{app:limits} how a sharp squamous to columnar transition can arise as the ratio $\Gamma_A/\Gamma_a$ varies when $a_0>2^{2/3}3^{7/6}\approx 5.71$ when $\Gamma_L=0$, mirroring a finding by \cite{sarkar2025} using a closely related model.  Fig.~\ref{fig:convol} below illustrates how, for $\Gamma_L> 0$, cells can become floppy through a rapid but continuous transition.}

\subsection{Regular hexagonal cells: external loading}

\begin{figure}
\centering
\includegraphics[width=\columnwidth]{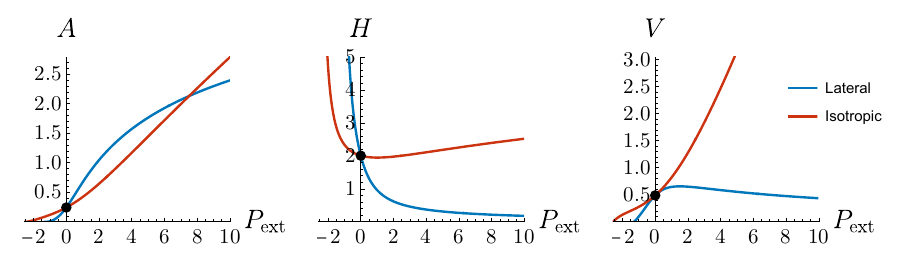}
\caption{Hexagonal cells in the baseline configuration (black dot) are perturbed by an external load $P_{\mathrm{ext}}$, that is compressive for $P_{\mathrm{ext}}<0$.  Lateral compression (blue) of a monolayer of hexagonal cells is modelled by (\ref{eq:lateralcomp}). Isotropic compression (red) of a hexagonal cell is modelled by (\ref{eq:isotropiccomp}). Changes in (a) area, (b) height and (c) volume are shown in response to lateral and isotropic loading.}
\label{fig:crowding}
\end{figure}

Cells in a monolayer that is laterally confined may experience compression due to crowding.  This is illustrated in Fig.~\ref{fig:crowding}(a), which shows (by solving (\ref{eq:lateralcomp})) how, for a cell experiencing uniform in-plane compression $P_{\mathrm{ext}}\mathsf{I}_\perp$ (with $P_{\mathrm{ext}}<0$), reduction of apical area and apical perimeter is balanced by a dramatic increase in cell height, even under a modest load, while maintaining roughly constant total surface area and volume.  A similar strongly anisotropic response arises in cells under isotropic compression with stress $P_{\mathrm{ext}}\mathsf{I}$ (with $P_{\mathrm{ext}}<0$; Fig.~\ref{fig:crowding}b), although in this instance the cell volume can be driven to zero under sufficient load (Fig.~\ref{fig:crowding}c).   

The slopes of the volume curves in Fig.~\ref{fig:crowding}(c) illustrate the cell bulk modulus under in-plane and isotropic compression.  The full stiffness matrix is evaluated in Appendix~\ref{app:stiffness}.  Second derivatives of the energy in (\ref{eq:fullstiff}) illustrate how the coupling between height, perimeter and area variations in the present 3D formulation contribute to stiffness.  The first derivatives of the energy in (\ref{eq:fullstiff}) illustrate the contribution of prestress.

\subsection{Disordered Monolayers}
\label{sec:dis}

\begin{figure*}
    \centering
    \includegraphics[width=0.75\linewidth]{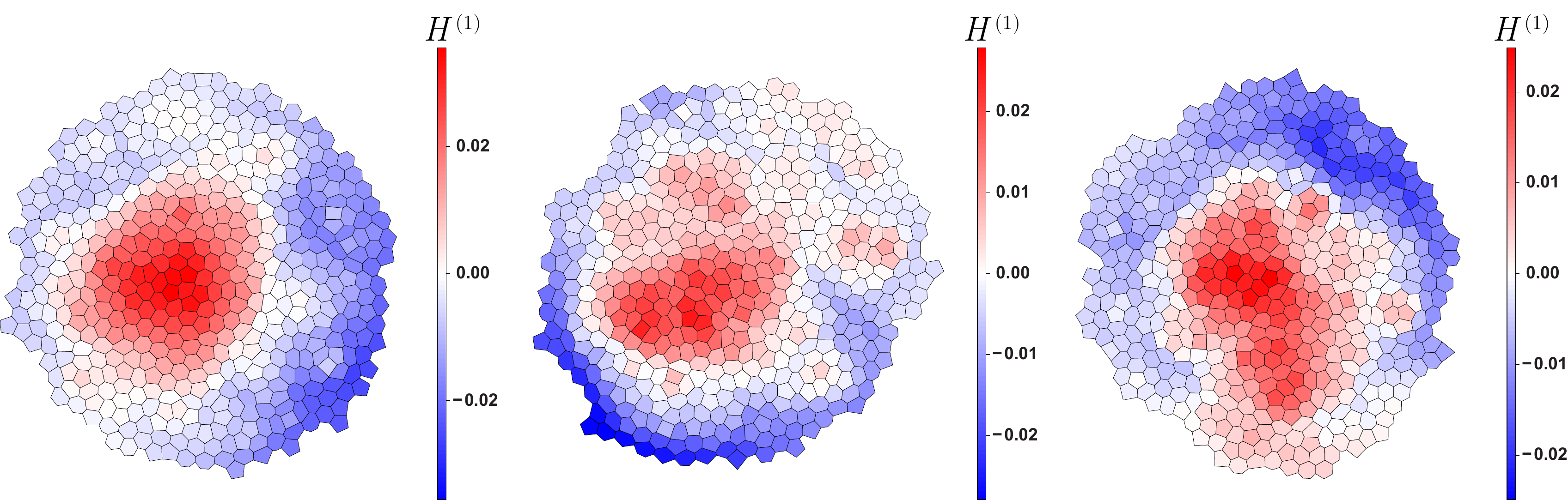}
    \caption{Distributions of  cell height perturbations $H^{(1)}$ evaluted using (\ref{eq:heightcorr}), for three different monolayers using baseline parameters, $L_0=2, \Gamma_a = 0.1, \Gamma_A = -0.5, \Gamma_L=0.1$ {\color{black}and $a_0=1$}.}
    \label{fig:heightvariation}
\end{figure*}

To understand better the impact of crowding on cell elongation, we grew disordered monolayers as described in Sec.~\ref{sec:config}, and then evaluated small-amplitude height variation across each monolayer (Fig.~\ref{fig:heightvariation}), adopting the model (\ref{eq:moden}) that penalises height variations between neighbouring cells.  Using baseline parameters, for which individual cells are rigid, realisations in Fig.~\ref{fig:heightvariation} show cell elongation nearer the centre of the monolayer, and contraction towards the periphery.  This is likely because divisions that take place during the growth of the monolayer promote crowding, increasing lateral forces on cells towards the centre of the monolayer and promoting elongation, as illustrated in Fig.~\ref{fig:crowding}. The Laplacian operator $\mathcal{L}$ in (\ref{eq:moden}), which averages over neighbouring cells, promotes a long-range response in height deflection; in this illustrative model, the stiffness of the height constraint (represented by the parameter $\Lambda$ in (\ref{eq:moden}) is independent of the monolayer's resistance to in-plane shear. 

Figure~\ref{fig:Vary_L0} shows how, for identical hexagons, increasing $L_0$ from its baseline value to approximately $2.83$ takes the system through a rigidity transition at which $P^{2D}$ and $T^{2D}$ fall to zero.  At this point, 8 of the 9 non-zero eigenvalues of the Hessian of the hexagonal state fall to zero (Fig.~\ref{fig:evals}), marking the emergence of floppy states with broken symmetry.  We now assess the comparable transition for an isolated disordered monolayer, grown as described in Sec.~\ref{sec:config}.

\begin{figure*}
    \centering
    \includegraphics[width=\linewidth]{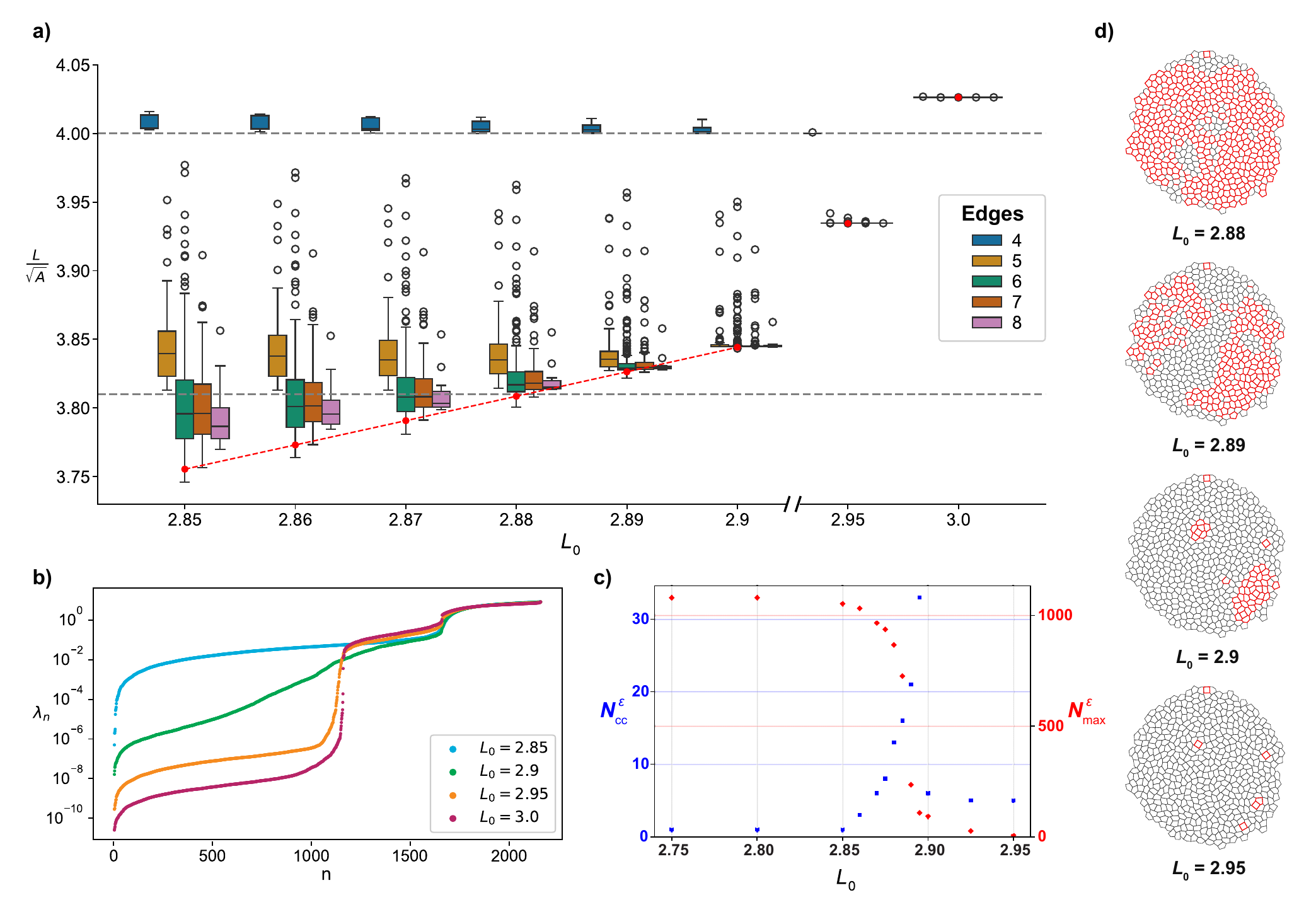}
    \caption{Simulations of a monolayer with $N_c=500$ cells and $N_v=1080$ apical vertices, successively relaxed through values of $L_0$.  (a) Distribution of the shape parameter ${L_h}/{\sqrt{A_h}}$ for varying $L_0$ with $\Gamma_a=0.1, \Gamma_A=-0.5, \Gamma_L=0.1$ {\color{black}and $a_0=1$}, coloured by the number of apical edges of each cell. White circles indicate outliers and the whiskers show the data extent excluding outliers, while the box shows the interquartile range. The red dots indicate the value of the shape parameter for an array of hexagons, which here is above the value at transition of $s_6\approx 3.72$.  Thresholds $s_4=4$ and $s_5\approx 3.81$ (see (\ref{eq:isoperim})) are shown with dashed lines. (b) The eigenvalues of the Hessian for the monolayers show in (a), for $L_0=\{2.85, 2.9, 2.95, 3.0\}$. (c) Number of connected components $N_{cc}^\epsilon$ in each monolayer (blue squares, left axis) and number of vertices in the largest connected component $N^\varepsilon_{\max}$ (red diamonds, right axis) as $L_0$ is varied. (d) Monolayers for $L_0= \{2.88,2.89, 2.9, 2.95\}$ showing edges with tension above a threshold value of $\varepsilon=0.0005$ in red.}
    \label{fig:shapeparam_L0_transition}
\end{figure*}

Figure~\ref{fig:shapeparam_L0_transition}(a) shows distributions of the shape parameter $L_h/\sqrt{A_h}$ for increasing $L_0$.  For reference, the shape parameter of identical hexagons (satisfying (\ref{eq:flo})) is also shown; this passes through $s_6\approx 3.72$ at $L_0\approx 2.83$.  In contrast, for the disordered monolayer the empirical threshold $s_5\approx 3.81$ provides a closer approximation of the rigidity transition, consistent with prior studies of the 2D model \cite{bi2015}, although the transition is not sharp.  It is marked by narrowing of the distributions of shape parameter among cells with different numbers of sides, so that an increasing proportion assume their target apical area and target apical perimeter.  Despite the dramatic collapse in the shape-parameter distribution onto the value predicted by (\ref{eq:flo}) at $L_0\approx 2.89$, it is limited to the majority of cells with at least five sides, as a few four-sided-cells persist, as do a few cells with $n\geq 5$ sides that do not meet the isoperimetric threshold.  Even at $L_0=2.95$, a few square cells persist that do not yet meet the target area and perimeter constraints.  These provide islands of stiffness in a sea of floppy cells. 

Figure~\ref{fig:shapeparam_L0_transition} (b) shows the $2N_v$ eigenvalues of the Hessian (Appendix~\ref{app:dynamics}) for increasing values of $L_0$.  At $L_0=2.85$ the monolayer is rigid; the largest $N_c$ modes of the Hessian's spectrum are dilational \cite{cowley2024}.  In general, $2N_c$ of the $2N_v$ modes can be associated with material stiffness (\hbox{i.e.} modes that change $U$), with the remainder providing geometric stiffness via prestress \cite{damavandi2022a, damavandi2022b, tong2023, cowley2024}.  As $L_0$ increases, the eigenvalues of modes providing geometric stiffness collapse towards zero, leaving $N_c$ shear and $N_c$ dilational modes \cite{cowley2024}.  Despite having material modes with non-zero eigenvalues, the spectrum shows how the monolayer becomes floppy once $L_0\approx 3$, although the transition is gradual.  

The component of the Hessian providing material stiffness is associated with second derivatives of $U$; the remainder (the $\bar{U}_\mathsf{s}$ term in (\ref{eq:hess})) provides geometric stiffness via the pre-stress fields $P^{2D}$ and $T^{2D}$ of the equilibrium state.  The component of the Hessian relating to $T^{2D}$ involves a vertex Laplacian of the form $\mathsf{A}^{++\top} \mathsf{A}^{++}$, where $\mathsf{A}^{++}$ is the apical edge-vertex incidence matrix (Appendix~\ref{app:matrices}), with entries weighted by edge tensions.  This motivates consideration of a `pruned' network made from apical cell edges having tension above a threshold value $\varepsilon$, for some $0<\varepsilon\ll 1$.  Examples of such networks are illustrated in Fig.~\ref{fig:shapeparam_L0_transition}(d).  Replacing $T^{2D}$ (in a given monolayer realisation) with an indicator function in the weighted vertex Laplacian yields an $N_v\times N_v$ topological operator $\mathcal{L}^\varepsilon$ (\ref{eq:prulap}), which varies with $L_0$, having $\mathcal{N}^\varepsilon(L_0)$  zero eigenvalues, where $1\leq N^\varepsilon\leq N_v$. Let $N^\varepsilon_{v0}(L_0)$ be the number of vertices that are not connected to an edge of the pruned network and $N^\varepsilon_{cc}(L_0)$ be the number of connected components of the pruned network (containing at least one edge).  Then $\mathcal{N}^\varepsilon= N^\varepsilon_{v0}+ N^\varepsilon_{cc}$; there are $N^\varepsilon_{v0}$ rows of zeros in $\mathcal{L}^\varepsilon$, and each connected component has a single zero eigenmode.  The network at $L_0=2.89$, for example, has 21 connected components, ranging from a single edge to the largest containing $235$ vertices (Fig~\ref{fig:shapeparam_L0_transition}c).  As $L_0$ increases across the rigidity transition, $N^\varepsilon_{cc}$ rises to a peak (comparable in magnitude to $\sqrt{N_v}\approx 33$) before falling abruptly towards 5 (Fig.~\ref{fig:shapeparam_L0_transition}(c)).  This value corresponds to the five quadrilateral cells in the monolayer that remain stiff at $L_0=2.95$ (Fig.~\ref{fig:shapeparam_L0_transition}a,d).

As $\varepsilon$ is an arbitrary threshold, the precise value of $L_0$ at which $N_{cc}^\varepsilon$ reaches its peak will depend on $\varepsilon$.  However this construction demonstrates a more general point, namely that the growing number of eigenvalues of $\mathcal{H}$ below a threshold ($10^{-5}$, say, in Fig.~\ref{fig:rigidity transition}(b)), as $L_0$ increases, can be interpreted primarily as modes of individual vertices that become isolated as adjacent edges lose tension, supplemented with zeros associated with connected components of the fragmenting network.  The size of the largest connected component (as measured by the number of vertices) falls from $N_v$ towards zero over a range of $L_0$ values, (Fig.~\ref{fig:shapeparam_L0_transition}d), showing the gradual nature of the transition when the monolayer is disordered.

\section{Discussion}

We have shown how a planar epithelium, modelled using a mechanical energy $U$ that reflects the 3D structure of cells, can be represented in a 2D formulation.  To achieve this, we introduced a constraint (\ref{eq:moden}) that restricted height variations between neighbouring cells.  A global force balance was used to determine the height $H$ of the cell population.  This, together with the involvement of an energy that penalises deviations of the total cell area from a target value, leads to a 2D model that is distinct from the classical formulation in which apical area $A$ and perimeter $L$ contribute separately to $U$.  The coupling of $A$, $H$ and $L$ leads to a more intricately-structured parameter space (Figs~\ref{fig:rigidity transition}, {\color{black}\ref{fig:a0_rigidity}}) that reveals multiple pathways to rigidity transitions, regulated by variation of {\color{black}five} independent dimensionless constitutive parameters.  This suggests that a monolayer changing state, for example in a wound-healing response, need not rely wholly on the cortical actin structures at the cell edge. This is consistent with prior models \cite{merkel2018geometrically} demonstrating regulation of rigidity via variations of the bulk target area through, for example, the parameter $a_0^*/(V_0^*)^{2/3}$.  
{\color{black}Like \cite{sarkar2025}, our model predicts a sharp squamous to columnar transition arising when the cell volume is constrained, provided $a_0^*/(V_0^{*})^{2/3}$ is sufficiently large.  Our findings rest on an assumption of apical-basal symmetry, so are likely to be of less relevance to epithelia that adhere strongly to a basement membrane, which can be expected to promote monolayer rigidity \cite{rozman2024basolateral}.}

This 3D perspective enables us to reinterpret prior 2D predictions of cell stress.  Strictly speaking, the 2D model identifies stress resultants, integrated across the depth of the cell.  These are naturally characterised in terms of in-plane isotropic and deviatoric (shear) components.  In the 3D context, this description is expanded to incorporate bulk isotropic stress and an additional bulk shear stress (\ref{eq: Peff}, \ref{eq:stresscomponents}b) that is associated with height reduction balancing apical area expansion; for a cell in vertical equilibrium, $2/3$ of the in-plane isotropic stress contributes to the bulk isotropic stress, and the remaining $1/3$ contributes to the bulk shear stress.  (The present model assumes apical-basal symmetry, and therefore does not address the full set of shear deformations.)  The model accommodates (elongated) columnar cells, particularly if the lateral adhesion is strong (Fig.~\ref{fig:Vary_Gamma_A}), or the preferred apical perimeter is sufficiently small (Fig.~\ref{fig:Vary_L0}); the bulk shear stress in such cases is regulated by $P_h^{\mathrm{eff}, 2D}$ via (\ref{eq:strf}), highlighting the role of the apical face in determining the overall mechanical state of the cell.   

The 3D formulation also reveals the connection between the functional dependence of energy on apical area, apical perimeter and cell height and the stiffness of a cell experiencing a prescribed deformation (Appendix~\ref{app:stiffness}).  The material stiffness arises from non-zero second derivatives of $U$ with respect to $A$, $L$ and $H$: the full set of nine second derivatives contribute resistance to out-of-plane shear deformation in (\ref{eq:outofplane}).  Stiffness also arises from prestress (associated with first derivatives of $U$ in (\ref{eq:outofplane}), for example).  For in-plane deformations, (\ref{eq:ipsd}) shows how prestress associated with cortical tension is lost at a rigidity transition; this is illustrated by fragmentation of the networks shown in Fig.~\ref{fig:shapeparam_L0_transition}(d), leading to isolated patches of stiffness as rigidity is lost.  We characterised this by counting zero eigenvalues of a vertex Laplacian (\ref{eq:prulap}) and the connected components the associated graph.  An important caveat of the analysis in Appendix~\ref{app:stiffness} is that a monolayer of disordered cells in a rigid state can be expected to undergo non-affine deformations under an external load, and non-affine effects become exaggerated in the floppy state, as illustrated in Fig.~\ref{fig:2D_fig}.  Computations of deformations under simple loads (Fig.~\ref{fig:crowding}) illustrate the strongly nonlinear anisotropic response arising from the combined linear and quadratic functions in the chosen mechanical energy.  For the chosen energies, cells can become columnar under modest external compression (Fig.~\ref{fig:crowding}) or under modest parameter variation (Fig.~\ref{fig:rigidity transition}). 

We grew disordered monolayers by allowing a sequence of random cell divisions, followed by relaxation to the nearest equilibrium.  This generates in-plane stresses that, in the rigid regime, cannot be resolved via neighbour exchanges.  Our simulations show evidence of elevated in-plane compression towards the centre of the monolayer, that promotes local cell elongation (Fig.~\ref{fig:heightvariation}).  It will be helpful to assess these findings against 3D simulations that do not assume apical-basal symmetry, and which allow more fine-grained variations of height of the apical vertices.  

In summary, the present model shows how 3D bulk mechanical effects can be incorporated in a 2D modelling framework, even when accommodating an out-of-plane force balance.  The model has the flexibility to capture independent variations in cell aspect ratio and monolayer stiffness, to describe in-plane and bulk stresses, and it demonstrates how long-range cell height variations can arise in a growing tissue.

\bigskip
\hrule
\bigskip

\begin{acknowledgments}
This work was supported by The Leverhulme Trust (RPG-2021-394) and a Wellcome Trust Career Development Award (225408/Z/22/Z).  For the purpose of open access,  the authors have applied a Creative Commons Attribution (CCBY) licence to any Author Accepted Manuscript version arising.
\end{acknowledgments}

\appendix

\section{Incidence and adjacency matrices}
\label{app:matrices}

Assigning orientations to all quantities, signed incidence matrices $\mathsf{A}$, $\mathsf{B}$ and $\mathsf{C}$ with components $C_{hi}=\{\mathsf{C}\}_{hi}$, $B_{ij}=\{\mathsf{B}\}_{ij}$ and $A_{jk}=\{\mathsf{A}\}_{jk}$ define topological arrangements.  Specifically, an edge orientation is defined by a tangent with magnitude equal to the edge length; a face orientation is defined by a normal with magnitude equal to the face area; $C_{hi}=1$ ($-1$) if the normal to face $i$ points into (out of) cell $h$; $B_{ij}=1$ ($-1$) if edge $j$ bounds face $i$ and is congruent (or anticongruent) to its orientation in a right-handed sense; $A_{jk}=1$ ($-1$) if edge $j$ points into (out of) vertex $k$.  The components of $\mathsf{A}$, $\mathsf{B}$ and $\mathsf{C}$ are zero otherwise. 

$\mathsf{A}$ includes a sub-block $\mathsf{A}^{++}$ connecting $N_e^+$ apical edges to $N_v^+$ apical vertices; $\mathsf{B}$ has sub-blocks $\mathsf{B}^{++}$ connecting $N_f^+$ apical faces to $N_e^+$ apical edges and $\mathsf{B}^{l+}$ connecting $N_f^l$ lateral faces to $N_e^+$ apical edges; and $\mathsf{C}$ has a blocks $\mathsf{C}^+$ and $\mathsf{C}^l$ connecting  to $N_c$ cells to $N_f^+$ apical and $N_f^l$ lateral faces, respectively.  Thus
\begin{subequations}
\begin{align}
    \mathsf{A}&=\left(\begin{matrix}\mathsf{A}^{++} & \mathsf{0} \\ \mathsf{A}^{l+} & \mathsf{A}^{l-} \\ \mathsf{0} & \mathsf{A}^{--}\end{matrix}\right),\\
    \mathsf{B}&=\left(\begin{matrix}\mathsf{B}^{++} & \mathsf{0} & \mathsf{0} \\ \mathsf{B}^{l+} & \mathsf{B}^{ll} & \mathsf{B}^{l-} \\ \mathsf{0} & \mathsf{0} & \mathsf{B}^{--}\end{matrix}\right), \\
    \mathsf{C}&=\left(\begin{matrix}\mathsf{C}^{+} & \mathsf{C}^{l} & \mathsf{C}^{-} \end{matrix}\right).
\end{align}
\end{subequations}
 The requirements that $\mathsf{C}\mathsf{B}=\mathsf{0}$ and $\mathsf{B}\mathsf{A}=\mathsf{0}$, arising because the 3D network is bounded by closed faces, imply
\begin{subequations}
    \begin{align}
    \left(\begin{matrix}
    \mathsf{C}^+\mathsf{B}^{++}+\mathsf{C}^l\mathsf{B}^{l+} &
    \mathsf{C}^l\mathsf{B}^l &
    \mathsf{C}^-\mathsf{B}^{--}+\mathsf{C}^l\mathsf{B}^{l-} 
    \end{matrix}\right)&=\mathsf{0}, \\
    \left(\begin{matrix}
    \mathsf{B}^{++}\mathsf{A}^{++} & \mathsf{0} \\
    \mathsf{B}^{l+}\mathsf{A}^{++} +\mathsf{B}^{ll}\mathsf{A}^{l+} &
    \mathsf{B}^{l-}\mathsf{A}^{--} +\mathsf{B}^{ll}\mathsf{A}^{l-} \\
    \mathsf{0} & \mathsf{B}^{--}\mathsf{A}^{--} 
    \end{matrix}\right)&=\mathsf{0}. 
\end{align}
\end{subequations}
The apical matrices $\mathsf{C}^+\mathsf{B}^{++}$ and $\mathsf{A}^{++}$
are the 3D analogues of the  face-edge and edge-vertex incidence matrices used in prior 2D studies \cite{jensen2020, jensen2023, cowley2024}; they satisfy the requirement that $(\mathsf{C}^+\mathsf{B}^{++})\mathsf{A}^{++}=\mathsf{0}$.  A 2D topological cell Laplacian is then
\begin{equation}
\mathcal{L}\equiv -\mathsf{C}^+\mathsf{B}^{++}\mathsf{E} \mathsf{B}^{++\top} \mathsf{C}^{+\top}.   
\label{eq:lap}
\end{equation}
This is regularised, as in \cite{jensen2023}, by taking $\mathsf{E}$ to be a diagonal $N_e\times N_e$ matrix for which $E_{jj}=1$ unless $j$ labels a peripheral apical edge of a peripheral cell in an isolated monolayer, in which case $E_{jj}=0$.  This regularisation ensures that $\mathcal{L}\mathsf{1}_c=0$, where $\mathsf{1}_c=(1,1,\dots)$ is the chain identifying all cells in a monolayer.

The adjacency matrices connecting vertices, edges, faces and cells are 
\begin{subequations}
\label{eq:adj}
\begin{align}
D^{\mathrm{cv}}_{hk}&\equiv\tfrac{1}{6}{\textstyle{\sum_{i,j}}}\vert C_{hi}\vert \vert B_{ij} \vert \vert A_{jk}\vert, &D^{\mathrm{ev}}_{jk}&\equiv\vert A_{jk}\vert
\\
D^{\mathrm{ce}}_{hj}&\equiv\tfrac{1}{2}{\textstyle{\sum_{i}}}\vert C_{hi}\vert \vert B_{ij} \vert, & D^{\mathrm{fe}}_{ij}&\equiv\vert B_{ij} \vert, \\
D^{\mathrm{fv}}_{ik}&\equiv\tfrac{1}{2}{\textstyle{\sum_{j}}} \vert B_{ij} \vert \vert A_{jk}\vert, & D^{\mathrm{cf}}_{hi}&\equiv\vert C_{hi}\vert.
\end{align}
\end{subequations}
The factor $\tfrac{1}{6}$ in $\mathsf{D}^{\mathrm{cv}}$ accounts for the fact that each vertex $k$ of cell $h$ neighbours three edges of the cell and each edge bounds two faces of the cell.  For the prismatic cells illustrated Fig.~\ref{fig:schematic}, distinguishing apical ($+$), lateral $(l)$ and basal $(-)$ objects, the adjacency matrices inherit the structure of the incidence matrices via
\begin{subequations}
\begin{align}
    \mathsf{D}^{\mathrm{fv}}&=\left(\begin{matrix}\mathsf{D}^{\mathrm{fv}++} & \mathsf{0} \\ \mathsf{D}^{\mathrm{fv},l+} & \mathsf{D}^{\mathrm{fv},l-} \\ \mathsf{0} & \mathsf{D}^{\mathrm{fv}--}\end{matrix}\right), \\
    \mathsf{D}^{\mathrm{ce}}&=\left(\begin{matrix}\mathsf{D}^{\mathrm{ce}+} & \mathsf{D}^{\mathrm{ce},l} & \mathsf{D}^{\mathrm{ce}-}\end{matrix}\right), \\
        \mathsf{D}^{\mathrm{cv}}&=\left(\begin{matrix} \mathsf{D}^{\mathrm{cv}+} & \mathsf{D}^{\mathrm{cv}-}\end{matrix}\right).
\end{align}
\end{subequations}

\section{Geometric identities}
\label{app:geom}

Given vertex locations $\mathbf{r}_k^\pm$, we construct apical cell edge tangents and centroids using 
\begin{equation}
    \mathbf{t}^+_j={\textstyle \sum_j} A_{jk}^{++}\mathbf{r}_k^+, \quad
    \mathbf{c}^+_j=\tfrac{1}{2}{\textstyle \sum_j} \vert A_{jk}^{++}\vert \mathbf{r}_k^+
    \label{eq:edges}
\end{equation}
respectively.  The apical perimeter and apical area of cell $h$ are then, respectively, for $h=1,\dots, N_c$
\begin{subequations}
    \label{eq:la}    
\begin{align}
    L_h^+&=\textstyle{\sum_{i,j}} \vert C_{hi}^l \vert \vert B_{ij}^{l+} \vert \vert\mathbf{t}^+_j\vert, \\
    A_h^+&=\tfrac{1}{2}{\textstyle{\sum_{i,j}}} \vert C_{hi}^+\vert B_{ij}^{++} \mathbf{c}_j^+\cdot(\mathbf{t}_j^+\times \hat{\mathbf{n}}^+_{hi}),
\end{align}
\end{subequations}
where $\hat{\mathbf{n}}^+_{hi} \equiv \hat{\mathbf{z}}$; $\hat{\mathbf{n}}_{hi}$ is the unit outward normal to face $i$ of cell $h$.  The area (\ref{eq:la}b) mirrors results in \cite{jensen2020, cowley2024}.  To capture the manner in which the apical perimeter changes as a result of displacing an apical vertex we use the relations 
\begin{subequations}
    \label{eq:ids}
\begin{align}
    \frac{\partial t_i^+}{\partial \mathbf{r}_k^+}&={\textstyle{\sum_j}} \vert B_{ij}^{l+}\vert A_{jk}^{++} \hat{\mathbf{t}}_j^+, \\
    \frac{\partial L_h^+}{\partial \mathbf{r}_k^+}&={\textstyle{\sum_{i,j}}}\vert C_{hi}^l\vert \vert B_{ij}^{l+} \vert A_{jk}^{++}\hat{\mathbf{t}}_j^+ =
    {\textstyle{\sum_{i}}}\vert C_{hi}^l\vert  \frac{\partial t_i^+}{\partial \mathbf{r}_k^+},
\end{align}
\end{subequations}
where the hat denotes a unit vector.

\section{2D Model}\label{2D Appendix}

\begin{figure}
    \centering
    \includegraphics[width=\linewidth]{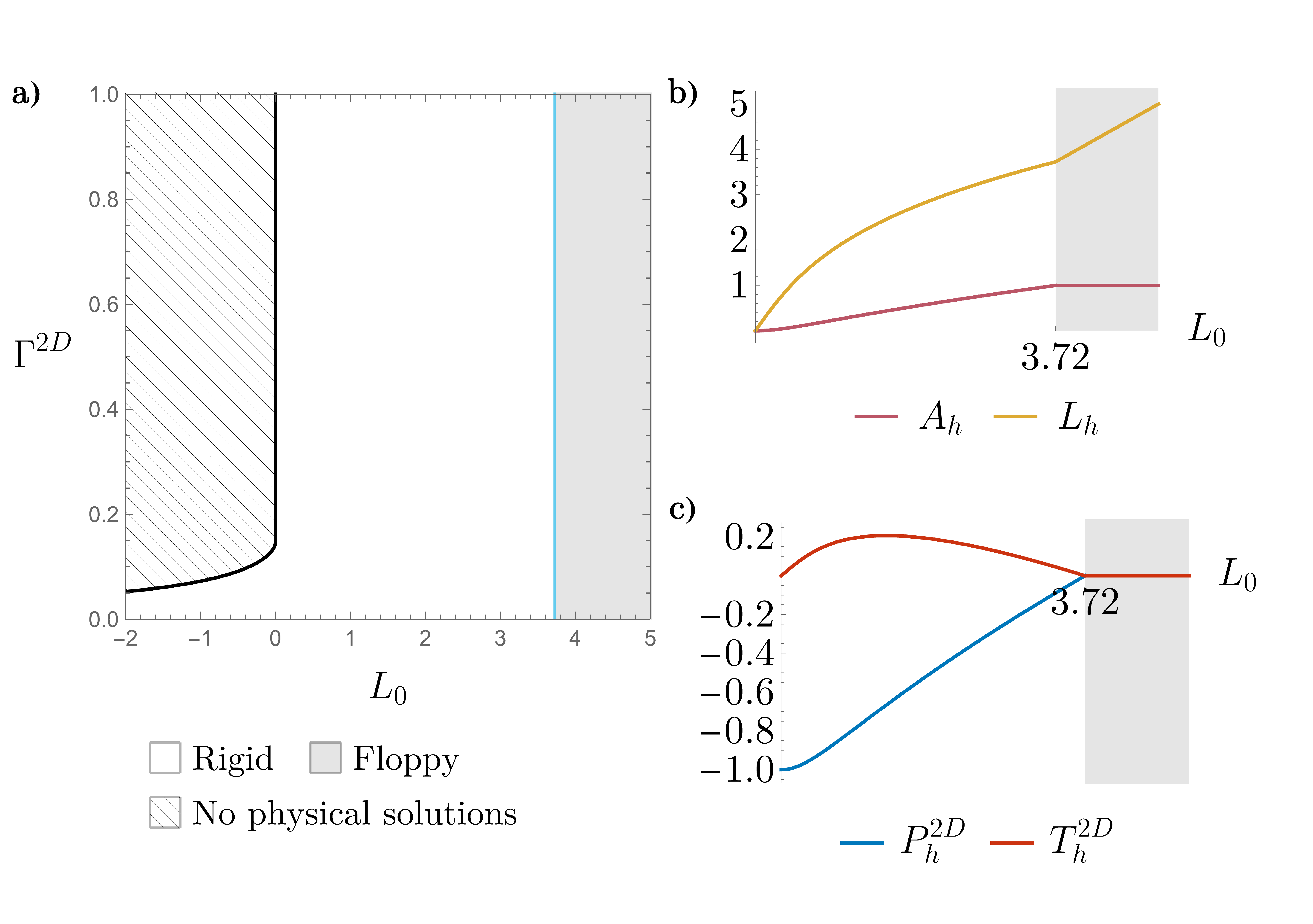}
    \caption{(a) The $(L_0,\Gamma^{2D})$-parameter map of the classical 2D vertex model \cite{farhadifar2007}.  Cells are rigid in the white region, becoming floppy for $L_0>s_6\approx 3.72$.  No solutions exist in the hatched region. (b) Area $A_h$ and perimeter $L_h$, and (c) pressure $P_h^{2D}$ and tension $T_h^{2D}$, are plotted as functions of $L_0$ for $\Gamma=$ 0.24; when $L_0=1$, $A_h, L_h$ take the baseline values in table \ref{tab:base_case_quatities}.}
    \label{fig:2D_fig}
\end{figure}

The classical 2D model is recovered from the present formulation by imposing $H=1$ in (\ref{eq:energy_AL}) (disregarding the vertical force balance), setting $\Gamma_a=0$ and defining $\Gamma^{2D}=2\Gamma_L$ and $L_0^{2D}=L_0-\Gamma_A/(4\Gamma_L)$.  This gives 
\begin{equation}
    U_h=\tfrac{1}{2}(A_h-1)^2+\tfrac{1}{2}\Gamma^{2D}(L_h-L_0^{2D})^2+\mathrm{constant}.
\end{equation}
The associated pressure and tensions are 
\begin{equation}
P_h^{2D}\equiv \frac{\partial U_h}{\partial A_h}=A_h-1, \quad T_h^{2D}\equiv \frac{\partial U_h}{\partial L_h}=\Gamma^{2D}(L_h-L_0^{2D}).
\end{equation}
Rigid hexagons satisfy (\ref{eq:hxf}a,b) {\color{black}(with $a_0=1$)}; floppy solutions satisfy $A_h=1$, $L_h=L_0^{2D}$, as illustrated in Fig.~\ref{fig:2D_fig}(b,c) (replicating \cite{farhadifar2007}, for example), which can be compared to figures~\ref{fig:Vary_L0}.  The parameter space map in Fig.~\ref{fig:2D_fig}(a) can be compared to Fig.~\ref{fig:rigidity transition}(c).  The rigid-floppy transition lies at $L_0^{2D}=s_6\approx 3.72$.  Rigid solutions collapse to zero area along $L_0^{2D}=0$ for $\Gamma^{2D}>\Gamma^*\equiv \sqrt{3}/12$.  Rigid solution branches terminate via a saddle-node bifurcation at $L_0^{2D}=-({8 \sqrt{2}}/(3 \Gamma ))\left(\Gamma^*-\Gamma^{2D} \right)^{\tfrac{3}{2}}$, with $0<\Gamma^{2D}\leq\Gamma^*$.

\section{Height variation across a constrained monolayer}
\label{sec:height}

In the main presentation of the model, we impose the requirement that $\mathsfbf{r}^\pm$ and $H$ evolve independently.  Alternatively, this constraint can be enforced by redefining the total energy to accommodate small variations in height between adjacent cells.  We write $\mathsf{H}=(H_1,\dots,H_{N_c})^\top$ and redefine $U$ as
\begin{equation}
\label{eq:moden}
    U={\textstyle\sum_h}\mathcal{U}(A_h,L_h, H_h)+\tfrac{1}{2} \Lambda \mathsf{H}^\top \mathcal{L}\mathsf{H}
\end{equation}
(see (\ref{eq:energy_AL})).  The Laplacian term (see (\ref{eq:lap})) mimics an external constraint that penalises height differences between neighbouring cells, when the constant $\Lambda$ is positive.  Then, noting that $\mathcal{L}=\mathcal{L}^\top$, the vertical force balance on individual cells $\partial U/\partial H_h=0$ becomes 
\begin{equation}
\mathsf{0}=\frac{\partial \mathcal{U}}{\partial{H}}(A_h,L_h,H_h)+\Lambda \left\{\mathcal{L}\mathsf{H} \right\}_h.
\end{equation}
Assuming $\Lambda\gg 1$ and writing $\mathsf{H}=\mathsf{H}^{(0)}+\mathsf{H}^{(1)}/\Lambda+\dots$, at leading order it follows that $\mathcal{L}\mathsf{H}^{(0)}=\mathsf{0}$.  This has a solution for which $\mathsf{H}^{(0)}$ is uniform across the monolayer, with $H^{(0)}_h=H^*$, say, for $h=1,2,\dots,N_c$.  At the following order, 
\begin{equation}
    \left\{\mathcal{L}\mathsf{H}^{(1)}\right\}_h=-\frac{\partial \mathcal{U}}{\partial {H}}(A_h,L_h,H^*).
\label{eq:heightcorr}
\end{equation}
As $\mathcal{L}\mathsf{1}=0$, so that $\mathsf{1}^\top \mathcal{L}=\mathsf{0}^\top$, it follows that a solvability condition for (\ref{eq:heightcorr}) is that $\sum_h (\partial \mathcal{U}/\partial H)(A_h,L_h,H^*)=0$, which is equivalent to the global force balance (\ref{eq:Ha0}).  We can then evaluate small corrections to cell height by inverting (\ref{eq:heightcorr}).  

\section{Asymptotic limits}
\label{app:limits}

Here we identify solutions of (\ref{eq:H}a,b), (\ref{eq:hxf}) describing configurations of rigid hexagonal cells, and floppy cells, in specific asymptotic limits.  

\textbf{Constrained total surface area.}  For $\Gamma_a\gg 1$, which constrains the total cell surface area to take the target value of unity, we reformulate the problem for rigid hexagons as (\ref{eq:hxf}a, b) with
\begin{subequations}
\label{eq:larga}
\begin{align}
0&=2A+LH-1, \\ P^{2D}&=H(AH-1)+2T, \\
0&=A(AH-1)+(T+\tfrac{1}{2}\Gamma_A)L, \\
T^{2D}&=TH+\tfrac{1}{2}\Gamma_A H +2\Gamma_L(L-L_0).
\end{align}
\end{subequations}
Here the tension $T$ is introduced as a Lagrange multiplier and (\ref{eq:larga}d) derives from (\ref{eq:H}c) using $T$ from (\ref{eq:PTs})$_2$.  The total area constraint (\ref{eq:larga}a) can be used to eliminate $H$; the vertical force balance (\ref{eq:larga}c) specifies $T$; and $P^{2D}$ and $T^{2D}$ are then combined in (\ref{eq:hxf}a, b) to find $l$.   Within this limit, columnar cells satisfy 
\begin{subequations}
\label{eq:elonga}
\begin{gather}
    L\approx \frac{1}{H}, ~~ A\approx \frac{\sqrt{3}{\color{black}a_0^3}}{24 H^2},~~ T\approx -\tfrac{1}{2}\Gamma_A+\frac{\sqrt{3}{\color{black}a_0^3}}{24H}, \\
    P^{2D}\approx -H, ~~ T^{2D}\approx \frac{\sqrt{3}{\color{black}a_0^3}}{12},\quad L_0\approx -\frac{\sqrt{3}{\color{black}a_0^3}}{48\Gamma_L},
 \end{gather}
 \end{subequations}
with $H\gg 1$.  Here the tension $T$ balances adhesive forces.  Floppy states under large $\Gamma_a$ are obtained by solving (\ref{eq:larga}) with $P^{2D}=0$, $T^{2D}=0$ for $A$, $H$, $L$ and $T$, {\color{black} which accommodates the solution for elongated cells
\begin{equation}
\label{eq:elong}
    A=0, \quad H=1/L_0, \quad L=L_0,\quad T=-\tfrac{1}{2}\Gamma_A,  
\end{equation}} 
shown on Fig.~\ref{fig:Vary_Gamma_a}(a,c).  Imposing (\ref{eq:s6}) on floppy states yields the large-$\Gamma_a$ limits shown in Fig.~\ref{fig:rigidity transition}(b,d,e) {\color{black}and \ref{fig:a0_rigidity}(a).}

\textbf{Constrained apical perimeter.} For $\Gamma_L\gg 1$, the apical perimeter is constrained to satisfy $L=L_0$ and we express the problem as
\begin{subequations}
\label{eq:largl}
\begin{align}
A&=(\sqrt{3}{\color{black}a_0^3}/24)L_0^2, \\ P^{2D}&=H(AH-1)+2\Gamma_a(2A+L_0 H-1), \\
0&=A(AH-1)+(\Gamma_a(2A+L_0H-1)+\tfrac{1}{2}\Gamma_A)L_0, \\ 
T^{2D}&=\Gamma_a(2A+ L_0H-1) H+\tfrac{1}{2}\Gamma_A H +T^c.
\end{align}
\end{subequations}
In this case, the hexagonal constraint (\ref{eq:largl}a) determines $A$, the vertical force balance (\ref{eq:largl}c) specifies $H$ and the Lagrange multiplier $T^c$ takes whatever value is needed to satisfy the force balance (\ref{eq:hxf}a, b).  This leads to asymptotes shown in Fig.~\ref{fig:Vary_Gamma_L}.  Floppy solutions satisfying (\ref{eq:flo}) in this limit are recovered by solving (\ref{eq:largl}b,c,d) with $P^{2D}=0$, $T^{2D}=0$ for $A$, $H$ and $T^c$ (and $L=L_0$).  Imposing (\ref{eq:s6}) in addition leads to limits plotted in Fig.~\ref{fig:rigidity transition}(c,d,f) {\color{black} and \ref{fig:a0_rigidity}(c)}.  Setting $H=0$ in (\ref{eq:largl}a,c) leads to 
\begin{equation}
\label{eq:sqlim}
    \Gamma_A=\frac{\sqrt{3}{\color{black}a_0^3}}{12}L_0+\Gamma_a\left(2-\frac{\sqrt{3}{\color{black}a_0^3}}{6}L_0^2\right),
\end{equation}
while (\ref{eq:largl}a,c) in the limit $H\rightarrow \infty$ requires
\begin{equation}
\label{eq:colim}
    \Gamma_a=-L_0^3{\color{black}a_0^6}/192.
\end{equation}

\textbf{Strong adhesion.}
The vertical force balance (\ref{eq:hxf}c) can be written as $0=A(AH-1)+(\Gamma_a(2A+LH-1)+\tfrac{1}{2}\Gamma_A)L$, and is then combined with (\ref{eq:H}b) to become $0=HA(AH-1)+L(T^{2D}-2\Gamma_L(L-L_0))$. Then, in the limit $\Gamma_A \rightarrow -\infty$, 
\begin{subequations}
   \label{eq:gaf}
\begin{align}
H&\approx \frac{2\Gamma_a^2-\Gamma_L}{2 \Gamma_a (\Gamma_a-\Gamma_L L_0)}\Gamma_A, \quad L \approx \frac{\Gamma_a-\Gamma_L L_0}{2\Gamma_a^2-\Gamma_L}, \\
A&\approx \frac{2 \Gamma_a \Gamma_L (2 \Gamma_a L_0-1) (\Gamma_a-\Gamma_L L_0)}{\Gamma_A\left(2 \Gamma_a^2-\Gamma_L \right)^2}.
\end{align}
\end{subequations}
This represents the dominant balance $0\approx H(AH-1)+2\Gamma_a LH$ in (\ref{eq:H}a), $0\approx \Gamma_a LH^2+\tfrac{1}{2}\Gamma_A H$ in (\ref{eq:H}b) and $AH(AH-1)-2\Gamma_L(L-L_0)=0$ in the vertical direction.  The tension balance is similar to the rigid limit (\ref{eq:gar}) but the deviation from the target volume is balanced in the pressure condition by tension associated with the area of lateral faces, and in the vertical force balance by peripheral apical tension.  At a rigidity transition, cells become hexagonal and the geometric constraint (\ref{eq:s6}) cannot be met unless 
\begin{equation}
L_0\approx \Gamma_a/\Gamma_L \quad (-\Gamma_A\rightarrow \infty).
\label{eq:floplimlargeGa}
\end{equation}
This limit is shown in Fig.~\ref{fig:rigidity transition}(a,e,f).  

Additionally, for $\Gamma_A\rightarrow -\infty$, analysis of (\ref{eq:H}a,b), (\ref{eq:hxf}) reveals the limit 
\begin{equation}
A\approx \frac{96\Gamma_a^2}{\sqrt{3}{\color{black}a_0^3} \Gamma_A^2}, \quad L\approx \frac{48\Gamma_a}{\sqrt{3}{\color{black}a_0^3}(-\Gamma_A)}, \quad H\approx \frac{\sqrt{3}{\color{black}a_0^3}\Gamma_A^2}{96\Gamma_a^2},
\label{eq:gar}
\end{equation}
satisfying the dominant balance $AH-1\approx 0$ in (\ref{eq:H}a), $\Gamma_aLH^2+\tfrac{1}{2}\Gamma_A H\approx 0$ in (\ref{eq:H}b) and (\ref{eq:hxf}c), with area and perimeter satisfying the isoperimetric constraint (\ref{eq:hxf}b); each cell is close to its target volume and an inward tension associated with total surface area, which is dominated by lateral faces, balances the outward tension due to adhesion.  

{\color{black}\textbf{Constrained volume.}  Here we consider 
\begin{subequations}
\label{eq:convol}
    \begin{align}
    H&=1/A, \\
    A&=\sqrt{3}a_0^3 L^2/24, \\ 
    0&=AP+[\Gamma_a(2A+LH-1)+\tfrac{1}{2}\Gamma_A]L, \\
    P^{2D}&=HP+2\Gamma_a (2A+LH-1), \\
    T^{2D}&=\Gamma_a(2A+LH-1)H+\tfrac{1}{2}\Gamma_A H+ 2\Gamma_L(L-L_0), \\
    0&=2AP^{2D}+LT^{2D}
\end{align}
\end{subequations}
in the limit $\Gamma_a\sim \Gamma_A\sim \Gamma_L\ll 1$.  We include $P$ as a Lagrange multiplier enforcing the volume constraint $AH=1$.  (Floppy states satisfy (\ref{eq:convol}a,c,d,e) with $P^{2D}=T^{2D}=0$ and (\ref{eq:isoperim}).)  Setting $\Gamma_L=0$, scaling $P$, $P^{2D}$ and $T^{2D}$ on $\Gamma_A$ and setting $\hat{\Gamma}_A=\Gamma_A/\Gamma_a$, (\ref{eq:convol}) reduces to 
\begin{equation}
\label{eq:twocube}
    (9l^3-2)(\sqrt{3}(9l^3+4)-3a_0 l)=3\hat{\Gamma}_A l a_0
\end{equation}
where $l=\frac{1}{6}a_0^2 L$.  The left-hand-side of (\ref{eq:twocube}) is a product of two cubics: the first has a single zero at $\ell=(2/9)^{1/3}$ in $\ell >0$; the second cubic (proportional to $2A+LH-1$) has two positive zeros for $a_0>2^{2/3} 3^{7/6}\approx 5.71$.  These zeros also first appear at $\ell=(2/9)^{1/3}$, implying that $\ell$ becomes a multivalued function of $\gamma$ for $a_0>2^{2/3}3^{7/6}$.  Floppy states for $\Gamma_L=0$ and $a_0>2^{2/3}3^{7/6}$ exist along $\hat{\Gamma}_A=0$, connecting roots of (\ref{eq:twocube}) for which $2A+LH=1$.  Thus, as predicted by \cite{sarkar2025} using a slightly different model, the squamous-to-columnar transition with $\Gamma_L=0$ is abrupt and takes place at $\hat{\Gamma}_A=0$.  

\begin{figure}
\includegraphics[width=\columnwidth]{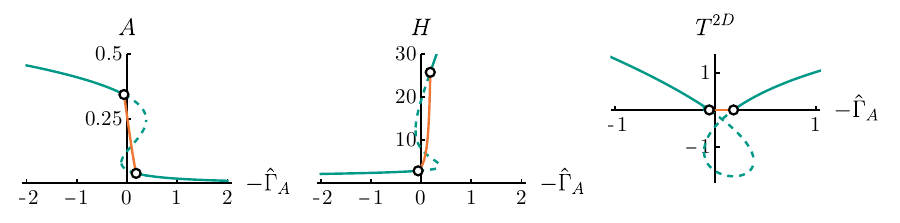}
    \caption{{\color{black}Example illustrating a sharp squamous-to-columnar transition when cell volume is constrained.  Green curves show solutions of (\ref{eq:convol}) for $a_0=7$, $\hat{\Gamma}_L\equiv \Gamma_L/\Gamma_a=0.1$ for varying $-\hat{\Gamma}_A\equiv -\Gamma_A/\Gamma_a$.  Orange lines show floppy states satisfying (\ref{eq:convol}a,c,d,e) with $P^{2D}=T^{2D}=0$ for which $A\leq \sqrt{3}a_0^3 L^2/24$.  Symbols indicate the two bifurcation points where the floppy and rigid solution branches intersect and rigid states lose stability (solid curves become dashed)}.}
    \label{fig:convol}
\end{figure}

For non-zero $\Gamma_L$, however, the branch of floppy solutions bends away from $\Gamma_A=0$ and intersects the branches of rigid solutions at new locations.  Fig.~\ref{fig:convol} illustrates one example.  The transition is continuous but a branch of floppy states connects the locally stable rigid squamous and columnar states.  

}

\section{Dynamics}
\label{app:dynamics}

\begin{figure*}    
    \centering
    \includegraphics[width=0.8\linewidth]{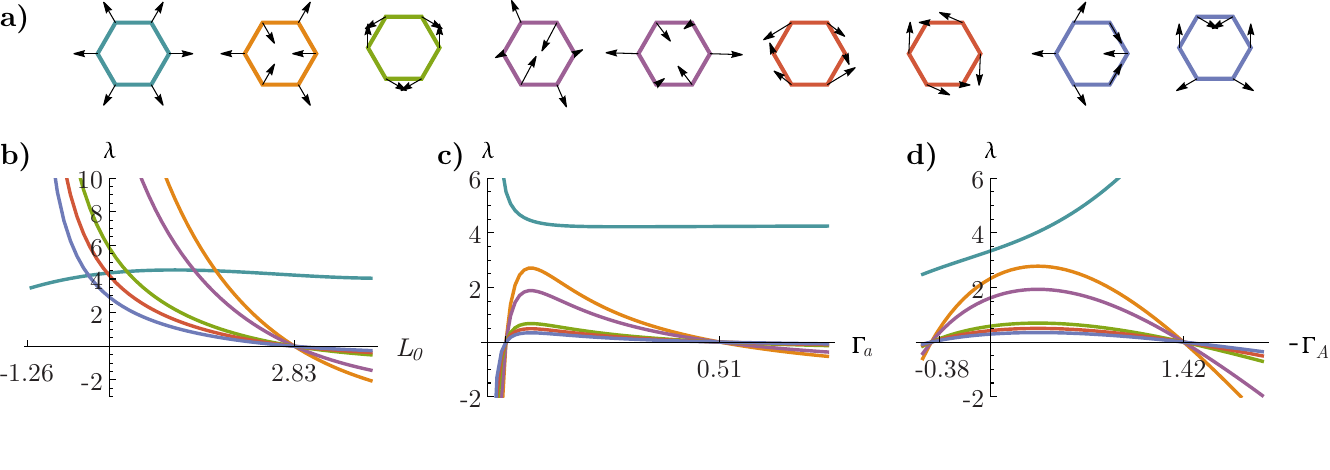}
    \caption{Eigenmodes (a) and eigenvalues of the Hessian $\mathcal{H}$ as (b) $L_0$, (c) $\Gamma_a$ and (d) $-\Gamma_A$ are varied, for a hexagon at baseline parameter values (Table~\ref{tab:base_case_quatities}). Eigenvalues correspond to the mode(s) of the same colour in (b--d); modes of the same colour in (a) share an eigenvalue.  The dilation mode is shown in cyan.  Translation and rotation modes (with zero eigenvalue) are not shown.}
    \label{fig:evals}
\end{figure*}

We consider dynamic evolution of a monolayer under the forces identified in (\ref{eq:p2t2}).  Following \cite{cowley2024}, we define $\mathsf{s}(\mathsf{r})=(A_1,\dots,A_{N_c},L_1,\dots,L_{N_c})^\top$ and collect energy derivatives as $U_{\mathsf{s}}\equiv (P_1^{2D},\dots,P_{N_c}^{2D}, T_1^{2D},\dots,T_{N_c}^{2D})^\top$. Imposing a drag on vertices, time evolution of  the system follows
\begin{equation}
\left(    \begin{matrix}
        1 & 0 \\ 0 & \mu 
    \end{matrix} \right) \frac{\mathrm{d}}{\mathrm{d}t} \left(    \begin{matrix}
        {\mathsfbf{r}} \\ {H} \end{matrix}\right) =-\left(\begin{matrix} {U}_\mathsf{s} {\mathsf{s}}_\mathsfbf{r}, \\ {U}_H \end{matrix}\right).
        \label{eq:fulldyn}
\end{equation}
Here the parameter $\mu$ models frictional resistance to height changes, relative to drag forces that resist in-plane vertex motion, and subscripts on $U$ denote derivatives. The energy can be expanded as 
\begin{multline}
U\approx \bar{U}+
(\bar{U}_\mathsf{s} \bar{\mathsf{s}}_\mathsfbf{r}, \bar{U}_H)
\left(\begin{matrix}\cdot \hat{\mathsfbf{r}}\\ \hat{H}\end{matrix}\right)\\+\tfrac{1}{2}
(\hat{\mathsfbf{r}}\cdot,\hat{H})\left( 
    \begin{matrix}
        \bar{\mathsf{s}}_{\mathsfbf{r}}^\top \bar{U}_{\mathsf{s}\mathsf{s}} \bar{\mathsf{s}}_\mathsfbf{r}+\bar{U}_{\mathsf{s}}^\top \bar{\mathsf{s}}_{\mathsfbf{r}\mathsfbf{r}} & \bar{\mathsf{s}}_{\mathsfbf{r}}^\top \bar{U}_{\mathsf{s}H}\\
        \bar{U}_{\mathsf{s}H}^\top \bar{\mathsf{s}}_{\mathsfbf{r}}  & \bar{U}_{HH}
    \end{matrix}
    \right)
    \left(\begin{matrix}\cdot \hat{\mathsfbf{r}} \\ \hat{H}\end{matrix}\right)+\dots
\end{multline}
A bar denotes evaluation at an equilibrium point (so that $\bar{U}_\mathsf{s}=\mathsf{0}$, $\bar{U}_H=0$) and hats denote small perturbations.  Provided $\bar{U}_{HH}\neq 0$, rapid relaxation of the depth (linearising (\ref{eq:fulldyn}) about the equilibrium and taking $\mu\rightarrow 0$) leads to the eigenvalue problem 
\begin{equation}
\lambda\hat{\mathsfbf{r}}=
\left[\bar{\mathsf{s}}_{\mathsfbf{r}}^\top (\bar{U}_{\mathsf{s}\mathsf{s}} -\bar{U}_{\mathsf{s}H} \bar{U}_{HH}^{-1}
        \bar{U}_{\mathsf{s}H}^\top)\bar{\mathsf{s}}_\mathsfbf{r}+\bar{U}_{\mathsf{s}}^\top \bar{\mathsf{s}}_{\mathsfbf{r}\mathsfbf{r}} 
    \right]
    \cdot \hat{\mathsfbf{r}},
    \label{eq:hess}
\end{equation}
where the solutions $\lambda$ are the decay rates of small perturbation.  The matrix operator in (\ref{eq:hess}) is the Hessian $\mathcal{H}$.  

Figure \ref{fig:evals} shows the values of the non-zero eigenvalues of $\mathcal{H}$ for an isolated hexagonal cell, for varying $L_0, \Gamma_a$ and $\Gamma_A$, as discussed in Sec.~\ref{sec:results}.  In all cases, all symmetry-breaking modes simultaneously become unstable when $T^{2D}=0$. The cell height vanishes at $\Gamma_A\approx 1.42$ when $\Gamma_A$ is varied.  All other unstable points correspond to a loss of in-plane rigidity. Eigenvalues of the symmetry-breaking modes diverge to $\pm \infty$ when the equilibrium cell area and perimeter go to zero.  

For some parameter values, multiple equilibrium solutions for rigid hexagonal cells can coexist; the Hessian in (\ref{eq:hess}) was used to confirm the stability of solutions plotted in Fig.~\ref{fig:rigidity transition}.

The component of the Hessian $\bar{U}_{\bar{\mathsf{s}}}^\top \mathsf{s}_{\mathsfbf{r}\mathsfbf{r}}$ in (\ref{eq:hess}) contains contributions from pre-stress through the vector $\bar{U}_{\mathsf{s}}$ of equilibrium pressures $\bar{\mathsf{P}}^{2D}$ and tensions $\bar{\mathsf{T}}^{2D}$.  $\bar{\mathsf{s}}_{\mathsfbf{r}\mathsfbf{r}}$ contains derivatives of (\ref{eq:ids}).  Focusing on components relating to $\bar{\mathsf{T}}^{2D}$, we define an indicator function satisfying 
\begin{equation}
\mathcal{I}_j^{\varepsilon}=\begin{cases}1 & \vert \sum_{h}D^{\mathrm{ce}}_{hj} \bar{T}_h^{2D}\vert >\varepsilon, \\ 0 &\mathrm{otherwise}, 
\end{cases} \quad (h=1,\dots,N_c),
\end{equation}
which identifies edges carrying tension above a threshold. Evaluation of the derivative $\partial^2 L_h^+/\partial \mathbf{r}_k^+ \partial \mathbf{r}_{k'}^+$ (see \cite{cowley2024}) and multiplication by $\mathcal{I}_h^{\varepsilon}$ leads to consideration of the operator 
\begin{equation}
\label{eq:prulap}
\left\{\mathcal{L^\varepsilon}\right\}_{k k'}\equiv
    \sum_{j} \mathcal{I}_j^{\varepsilon}A^{++}_{jk} A^{++}_{jk'}.
\end{equation}
This modification of the full apical vertex Laplacian $\mathsf{A}^{++\top}\mathsf{A}^{++}$ retains contributions from edges of cells carrying  non-vanishing tension.

\section{Stress}
\label{app:stress}

The stress for the symmetric cells considered here can be derived directly by using the energy (\ref{eq:energy_AL}) and writing the force on vertex $k$ as  
\begin{align}
\mathbf{f}_{hk}=&\frac{\partial U}{\partial A_h}\frac{\partial A_h}{\partial \mathbf{R}_k}+ \frac{\partial U}{\partial L_h^+}\frac{\partial L_h^+}{\partial \mathbf{R}_k}+\frac{\partial U}{\partial L_h^-}\frac{\partial L_h^-}{\partial \mathbf{R}_k}+\frac{1}{Z_h}\frac{\partial U}{\partial H}{}\frac{\partial H}{\partial \mathbf{R}_k}\nonumber \\
=&\left(H\frac{\partial U}{\partial V_h}+2\frac{\partial U}{\partial a_h}\right) \frac{\partial A_h}{\partial \mathbf{R}_k}\nonumber\\
 +& \left (H \frac{\partial U}{\partial a_h}+ T_h^c+\tfrac{1}{2}\Gamma_A H\right )\frac{\partial L_h}{\partial \mathbf{R}_k}
+\frac{1}{Z_h}\frac{\partial U}{\partial H}\frac{\partial H}{\partial \mathbf{R}_k}, 
\end{align}
where ${\partial U}/{\partial H}=A_h \partial U /\partial {V_h}+L_h \partial U/\partial {a_h}+\tfrac{1}{2}\Gamma_A L_h$.  Here $Z_h$ is the number of apical vertices, across which the vertical force is distrubuted.  We then exploit 2D results \cite{nestor2018}, plus the observation that $\mathbf{R}_k\otimes (\partial H/\partial\mathbf{R}_k)=H\hat{\mathbf{z}}\otimes\hat{\mathbf{z}}$ when we enforce $\mathbf{R}_k=\mathbf{r}_k$ for vertices in the plane $z=0$, and $\mathbf{R}_k=\mathbf{r}_k+H\hat{\mathbf{z}}$ for cells in the plane $z=H$, to derive (\ref{eq:Vhsigma_3Da}).

\section{Stiffness}
\label{app:stiffness}

Consider the energy $\mathcal{U}(A,L,H)$ for a single cell (\ref{eq:energy_AL}).  In this Appendix, $h$ suffices are dropped and subscripts on $\mathcal{U}$ denote derivatives. 2D and 3D stresses (\ref{eq:Vhsigma_3Da}) can be written
\begin{equation}
V\boldsymbol{\sigma}=A\boldsymbol{\sigma}^{2D}+H \mathcal{U}_H \hat{\mathbf{z}}\otimes\hat{\mathbf{z}},\quad A\boldsymbol{\sigma}^{2D} = A\mathcal{U}_A\mathsf{I}_\perp + L\mathcal{U}_L\mathsf{Q}.
\label{eq:strs}
\end{equation}
Recall that $\mathsf{I}_\perp:\mathsf{Q}=1$, $2AP^{\mathrm{eff},2D}=2A\mathcal{U}_A+L\mathcal{U}_L$ and $3VP^{\mathrm{eff}}=2A\mathcal{U}_A+L\mathcal{U}_L+H\mathcal{U}_H$.  For a symmetric hexagonal cell, $\mathsf{Q}=\tfrac{1}{2}\mathsf{I}_\perp$.  We then impose an affine deformation $\mathsf{E}=\mathsf{E}_\perp + e \hat{\mathbf{z}}\otimes\hat{\mathbf{z}}$.  For example, isotropic expansion has $\mathsf{E}_\perp=e\mathsf{I}$ so that $\mathrm{Tr}(\mathsf{E})=3e$;  in-plane shear deformation has $e=0$ with $\mathsf{I}_\perp:\mathsf{E}_\perp=0$;  out-of-plane shear deformation has $\mathsf{E}_\perp=-\tfrac{1}{2}e\mathsf{I}_\perp$.  We now seek the stress generated by an affine displacement; this approach complements that of \cite{kim2024mean}, who evaluated the elastic modulus of a cell modelled in 3D as a truncated octahedron.

Under an arbitrary small deformation, perturbations in $A$, $L$, $H$ and $\mathsf{Q}$ are $\Delta A=A\mathsf{I}_\perp:\mathsf{E}$, $\Delta L=L\mathsf{Q}:\mathsf{E}$, $\Delta H=He$ and $\Delta\mathsf{Q}=\mathsf{B}:\mathsf{E}$ respectively, where the fourth-order tensor $\mathsf{B}$ (defined in \cite{nestor2018a}) satisfies $\mathsf{B}:\mathsf{I}_\perp=\mathsf{Q}$ and $\mathrm{Tr}(\mathsf{B}:\mathsf{E})=\mathsf{Q}:\mathsf{E}$.  Perturbing all quantities in (\ref{eq:strs}) and linearising leads to 
\begin{subequations}
\label{eq:fullstiff}
\begin{align}
V\Delta\boldsymbol{\sigma}&=\boldsymbol{\alpha}(\mathsf{I}_\perp:\mathsf{E}_\perp)+\boldsymbol{\beta}e+\boldsymbol{\gamma}\mathsf{Q}:\mathsf{E}_\perp+\delta \mathsf{B}:\mathsf{E}_\perp, \\
\boldsymbol{\alpha}=& A^2 \mathcal{U}_{AA}\mathsf{I}_\perp+(LA\mathcal{U}_{LA}-L\mathcal{U}_L)\mathsf{Q}\nonumber \\&\quad+(AH\mathcal{U}_{AH}-H\mathcal{U}_H)\hat{\mathbf{z}}\otimes\hat{\mathbf{z}},\\
\boldsymbol{\beta} =& (AH \mathcal{U}_{AH}-A\mathcal{U}_A)\mathsf{I}_\perp+(LH\mathcal{U}_{LH}-L\mathcal{U}_L)\mathsf{Q} \nonumber\\ &\quad +H^2 \mathcal{U}_{HH}\hat{\mathbf{z}}\otimes\hat{\mathbf{z}},\\
\boldsymbol{\gamma} =& AL\mathcal{U}_{AL}\mathsf{I}_\perp+(L^2 \mathcal{U}_{LL}+L\mathcal{U}_L)\mathsf{Q}+HL\mathcal{U}_{HL} \hat{\mathbf{z}}\otimes\hat{\mathbf{z}},\\
\delta =&L\mathcal{U}_L.
\end{align}
\end{subequations}

These coefficients contain second derivatives of the energy, contributing to material stiffness, plus first derivatives, contributing to geometric stiffness.  The first derivatives can be identified as prestresses in the baseline state in (\ref{eq:strs}).  Eq.~(\ref{eq:fullstiff}) demonstrates how the cross-coupling in $\mathcal{U}$ leading to non-zero mixed second derivatives can influence the cell stiffness.

The stress-strain relationship under isotropic expansion is then
\begin{subequations}
\label{eq:outofplane}
\begin{align}
V\Delta\boldsymbol{\sigma}&=\left[2\boldsymbol{\alpha}+\boldsymbol{\beta}+\boldsymbol{\gamma}+\delta \mathsf{Q}\right]e  \nonumber \\
&=\big\{(2A^2\mathcal{U}_{AA}+AL\mathcal{U}_{AL}+AH\mathcal{U}_{AH}-A\mathcal{U}_A)\mathsf{I}_\perp \nonumber \\
& \qquad +(2LA\mathcal{U}_{AL}+L^2\mathcal{U}_{LL}+LH\mathcal{U}_{HL}-L\mathcal{U}_L)\mathsf{Q} \nonumber \\
+(2&AH\mathcal{U}_{AH}+HL\mathcal{U}_{HL}+H^2\mathcal{U}_{HH}-2H\mathcal{U}_H)\hat{\mathbf{z}}\otimes\hat{\mathbf{z}}\big\}e.
\end{align}
\end{subequations}
For a hexagonal cell, this simplifies to 
\begin{subequations}
\label{eq:isoexp}
\begin{align}
V\Delta\boldsymbol{\sigma}&=\left[2\boldsymbol{\alpha}+\boldsymbol{\beta}+\boldsymbol{\gamma}+\delta \mathsf{Q}\right]e  \\
&=\big\{(2A^2\mathcal{U}_{AA}+2AL\mathcal{U}_{AL}+AH\mathcal{U}_{AH}-A\mathcal{U}_A \nonumber\\ &\qquad +\tfrac{1}{2}(L^2\mathcal{U}_{LL}+LH\mathcal{U}_{HL}-L\mathcal{U}_L))\mathsf{I}_\perp \nonumber \\
+(2&AH\mathcal{U}_{AH}+HL\mathcal{U}_{HL}+H^2\mathcal{U}_{HH}-2H\mathcal{U}_H)\hat{\mathbf{z}}\otimes\hat{\mathbf{z}}\big\}e.
\end{align}
\end{subequations}
The trace of (\ref{eq:isoexp}) recovers a bulk modulus, illustrated more directly in Fig.~\ref{fig:heightvariation}(c).  The anisotropic response to isotropic forcing, evident in Fig.~\ref{fig:heightvariation}, is consistent with different in-plane and out-of-plane stress responses in (\ref{eq:isoexp}).

In-plane shear deformation leads to 
\begin{align}
\label{eq:ipsd}
V\Delta\boldsymbol{\sigma}&=\big[AL\mathcal{U}_{AL}\mathsf{I}_\perp+(L^2 \mathcal{U}_{LL}+L\mathcal{U}_L)\mathsf{Q}\nonumber \\ & \quad +HL\mathcal{U}_{HL} \hat{\mathbf{z}}\otimes\hat{\mathbf{z}}\big]
\mathsf{Q}:\mathsf{E}_\perp+L\mathcal{U}_L \mathsf{B}:\mathsf{E}_\perp.
\end{align}
Two terms proportional to $L\mathcal{U}_L$ rely on cortical tension to generate stiffness.  In contrast, for the classical 2D model with $\mathcal{U}=\tfrac{1}{2}(A-1)^2+\tfrac{1}{2}\Gamma(L-L_0)^2$ and $T\equiv \mathcal{U}_L$, (\ref{eq:ipsd}) simplifies to 
\begin{equation}
V\Delta\boldsymbol{\sigma}=\Gamma L^2 \mathsf{Q}(\mathsf{Q}:\mathsf{E}_\perp) +L T \left[\mathsf{Q}(\mathsf{Q}:\mathsf{E}_\perp)+ \mathsf{B}:\mathsf{E}_\perp\right].
\end{equation}
The first term provides material stiffness (via $\mathcal{U}_{LL}=\Gamma$), the second geometric stiffness (via $U_L=T$).  In (\ref{eq:ipsd}), additional material stiffness arises from $U_{AL}$ and $U_{HL}$.

Finally, out-of-plane shear deformation leads to
\begin{align}
V\Delta\boldsymbol{\sigma}&=\left[-\boldsymbol{\alpha}+\boldsymbol{\beta}-\tfrac{1}{2}\boldsymbol{\gamma}-\tfrac{1}{2}\delta \mathsf{Q}\right]e \nonumber \\
&=\big\{(-A^2\mathcal{U}_{AA}-\tfrac{1}{2}AL\mathcal{U}_{AL}+AH\mathcal{U}_{AH}-A\mathcal{U}_A)\mathsf{I}_\perp \nonumber \\
& \qquad +(-LA\mathcal{U}_{AL}-\tfrac{1}{2}L^2\mathcal{U}_{LL}+LH\mathcal{U}_{HL}-L\mathcal{U}_L)\mathsf{Q} \nonumber \\
 +(-&AH\mathcal{U}_{AH}-\tfrac{1}{2}HL\mathcal{U}_{HL}+H^2\mathcal{U}_{HH}+H\mathcal{U}_H)\hat{\mathbf{z}}\otimes\hat{\mathbf{z}}\big\}e.
\end{align}
For a hexagonal cell, this simplifies to
\begin{align}
V\Delta\boldsymbol{\sigma}&=\left[-\boldsymbol{\alpha}+\boldsymbol{\beta}-\tfrac{1}{2}\boldsymbol{\gamma}-\tfrac{1}{2}\delta \mathsf{Q}\right]e \nonumber \\
&=\big\{(-A^2\mathcal{U}_{AA}-AL\mathcal{U}_{AL}+AH\mathcal{U}_{AH}-A\mathcal{U}_A \nonumber\\&\qquad-\tfrac{1}{4}L^2\mathcal{U}_{LL}+\tfrac{1}{2}LH\mathcal{U}_{HL}-\tfrac{1}{2}L\mathcal{U}_L)\mathsf{I}_\perp \nonumber \\
 +(-&AH\mathcal{U}_{AH}-\tfrac{1}{2}HL\mathcal{U}_{HL}+H^2\mathcal{U}_{HH}+H\mathcal{U}_H)\hat{\mathbf{z}}\otimes\hat{\mathbf{z}}\big\}e,
\end{align}
showing how all second derivatives of $U$ contribute to the material stiffness.

{\color{black}

\begin{figure}
    \centering
    \includegraphics[width=1\columnwidth]{a0_plots_tilde.png}
    \caption{{\color{black}Impact of varying $a_0$, presented following the format of Fig.~\ref{fig:Vary_L0}, except that geometric variables are plotted without implementing the rescaling in (\ref{eq:scalea0}).}  }
    \label{fig:vary_a0}
\end{figure}

\section{Target surface area}
\label{app:tsa}

The effects of varying the parameter $a_0$ are shown in Figs~\ref{fig:vary_a0} and \ref{fig:a0_rigidity}. Both figures show variables and parameters without implementing the scaling by $a_0$ in \eqref{eq:scalea0}. Axes and legend labels are given in terms of the scaled values used elsewhere but with the tilde explicitly shown here.

\begin{figure}
    \centering
    \includegraphics[width=1\columnwidth]{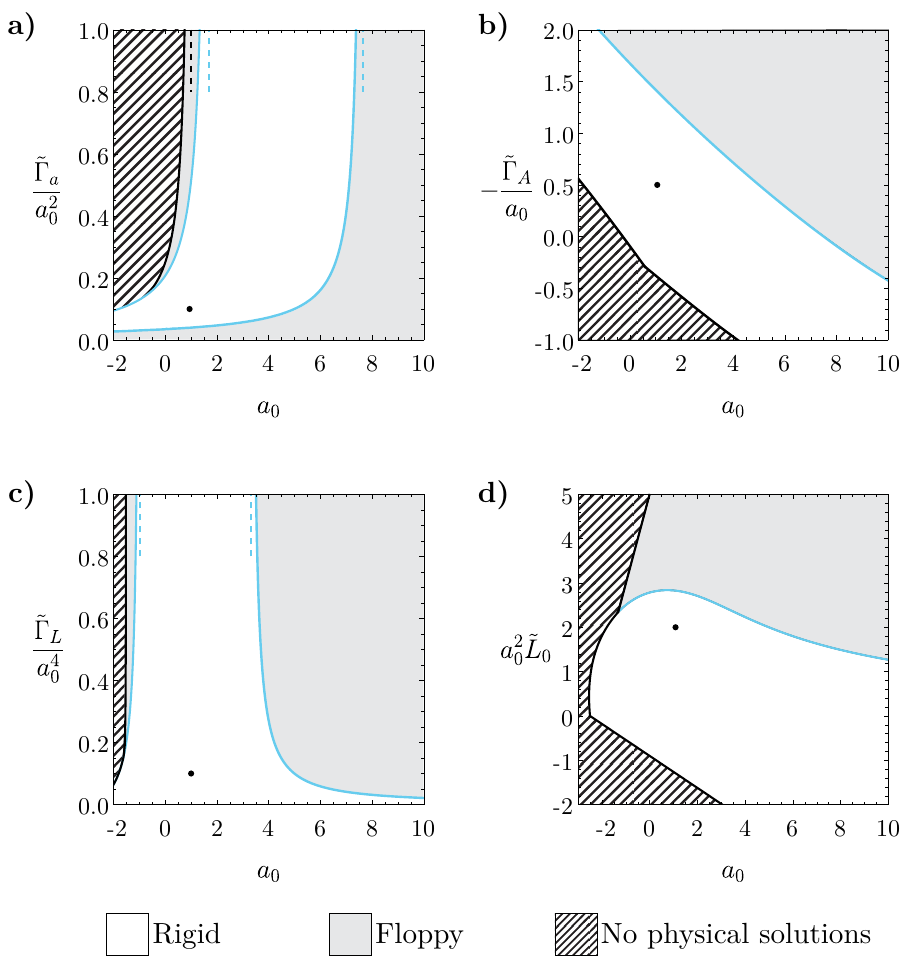}
    \caption{ {\color{black}Cross-sections of $a_0$ with $(L_0, \Gamma_a, -\Gamma_A, \Gamma_L)$-parameter space, intersecting the baseline case (black circle).  Parameters are plotted without implementing the rescaling in (\ref{eq:scalea0}).  In each plot, only two parameters are varied.  Hatching shows regions where there are no physical solutions. Grey regions show the floppy regime.  Blue solid lines show where $T^{2D}=0$, corresponding to a rigidity transition.  Blue dashed lines show asymptotes (\ref{eq:larga}) and (\ref{eq:s6}) with $P^{2D}=T^{2D}=0$ in (a), and (\ref{eq:largl}) with $P^{2D}=T^{2D}=0$ in (c).  The black dashed line in (a) shows (\ref{eq:elong}).}}
    \label{fig:a0_rigidity}
\end{figure}

}
\bibliography{refs}

\end{document}